\documentclass[seceq]{ptptex}

\usepackage{graphicx}
\usepackage{wrapft}

\newcommand{\g}{$\gamma$}
\newcommand{\msun}{{{\rm M}_{\odot}}}
\newcommand{\xte}{{\it RXTE}}
\newcommand{\ginga}{{\it Ginga}}

\newcommand{\gro}{{\it CGRO}}
\newcommand{\glast}{{\it GLAST}}
\newcommand{\sax}{{\it Beppo\-SAX}}
\newcommand{\ledd}{L_{\rm E}}
\newcommand{\ee}{$e^\pm$}

\newbox\grsign \setbox\grsign=\hbox{$>$} \newdimen\grdimen \grdimen=\ht\grsign
\newbox\simlessbox \newbox\simgreatbox \newbox\simpropbox
\setbox\simgreatbox=\hbox{\raise.5ex\hbox{$>$}\llap
     {\lower.5ex\hbox{$\sim$}}}\ht1=\grdimen\dp1=0pt
\setbox\simlessbox=\hbox{\raise.5ex\hbox{$<$}\llap
     {\lower.5ex\hbox{$\sim$}}}\ht2=\grdimen\dp2=0pt
\setbox\simpropbox=\hbox{\raise.5ex\hbox{$\propto$}\llap
     {\lower.5ex\hbox{$\sim$}}}\ht2=\grdimen\dp2=0pt
\def\simgreat{\mathrel{\copy\simgreatbox}}
\def\simless{\mathrel{\copy\simlessbox}}

\pubinfo{Supplement No.\ 155, 2004}
\notypesetlogo                       %comment in if to eliminate PTPTeX
\title{Radiative Processes, Spectral States and Variability of Black-Hole Binaries}

\author{Andrzej A. \textsc{Zdziarski}$^{1,}$\footnote{E-mail: aaz@camk.edu.pl} and Marek \textsc{Gierli\'nski}$^{2,3,}$\footnote{E-mail: Marek.Gierlinski@durham.ac.uk}
}

\inst{$^1$Centrum Astronomiczne im.\ Miko{\l}aja Kopernika, Bartycka 18, 00-716 Warszawa, Poland\\
$^2$Department of Physics, University of Durham, Durham DH1~3LE, UK\\
$^3$Obserwatorium Astronomiczne Uniwersytetu Jagiello\'nskiego, Orla 171, 30-244 Krak{\'o}w, Poland\\
}

\abst{ We review radiative processes responsible for X-ray emission in hard
(low) and soft (high) spectral states of black-hole binaries. The main process
in the hard state appears to be scattering of blackbody photons from a cold disk
by thermal electrons in a hot inner flow, and any contribution from nonthermal
synchrotron emission is at most small. In the soft states, blackbody disk
emission dominates energetically, and its high-energy tail is due to scattering
by hybrid, thermal/nonthermal electrons, probably in active regions above the
disk surface. State transitions appear to correspond to a variable inner radius
of the cold disk driven by changes of the accretion rate. The existence of two
accretion solutions, hot and cold, in a range of the accretion rate leads to
hysteresis in low-mass X-ray binaries. }

\begin{document}

\maketitle

\section{Introduction}
\label{intro}

We present here a brief review of energy and power spectra, radiative processes
giving rise to them, spectral states, and some aspects of variability of
black-hole binaries. Our emphasis is on interpretation of the X-ray and soft
\g-ray (hereafter X\g) observations in terms of simple physical models and
unifying various aspects of the wealth of the present observational data. We
also discuss millisecond flares recently discovered from Cyg X-1\cite{gz03},
where we present some important new results. We refer the reader to
Ref.~\citen{mr04} for an exhaustive review of observational aspects of
black-hole binaries, and, e.g., to Refs.\ \citen{d02,z99,b04} and \citen{z00} for some other relevant reviews.

In our presentation, we use the so-called $\nu F_\nu$ representation for both
energy and power spectra. The motivation for this choice is provided by a
trivial mathematic consideration. Namely, any differential dependence, e.g.,
flux per unit energy, ${\rm d}F/{\rm d}E$, equivalently expressed as
$F(E)$ or $F_E$, can be plotted as $F(E)$ vs $E$. However, when we plot $F$
vs.\ $\log E$, it should be $F(\log E)={\rm d}F/{\rm d}\log E$, i.e., $F(E)
({\rm d}\,\log E/{\rm d}E)^{-1}\propto E F(E)$. This gives us $F$ per a
logarithmic interval of $E$. Graphically, $EF(E){\rm d}\log E$ represents the
area under the curve, $F$, in a plot vs $\log E$. Then, a peak in this
representation shows us at which photon energy range most of the power is
radiated, or which frequency range corresponds to most of variability of a
source. An added benefit of this representation is that plots vs.\ photon energy
or wavelength become identical, as $\lambda F(\lambda) = E F(E)$. We also plot
the energy spectra with the same length per decade on each axis, in order to
enable direct comparison of the shape spectra on different figures.

On the other hand, it is common in X-ray astronomy to plot photon flux vs.\
energy. However, this makes it incompatible with most of other branches of
astronomy (e.g., those from the radio to the UV), where energy flux is
used. Also, energy is a much more fundamental quantity in physics than photon
number, with many physical laws concerning the former, not the latter (as
photon number is not conserved in most of physical processes). Indeed, all
quantities of radiative transfer, e.g., flux, specific and average intensity,
radiation pressure, are defined in terms of energy.

A practical consequence of plotting photon number per unit energy on a
logarithmic plot is that such spectra look usually extremely steep, covering
many orders of magnitude on the vertical axis, appear very similar to each
other, and make it very difficult to actually see any spectral features. This
is due to the artificial steepening of such a spectrum by $E^{-2}$, described
above.

It is also very common in X\g\ astronomy to plot instrumental counts instead of
photons or energy. The motivation for it is that high-energy instruments have
rather limited energy resolution, which makes the process of deconvolution not
unique. However, this purist point of view would actually require plotting only
instrumental counts, as plotting together predictions of a model and/or
residuals is already model-dependent. In fact, residuals given as, e.g., ratios
or contributions to $\chi^2$, are {\it identical\/} for either counts vs.\
model or a deconvolved spectrum vs.\ model. However, the latter has the
enormous advantages of presenting a physical spectrum, allowing comparison with
other spectra, and showing features originating in the source. The count-rate
plot, though model-independent, is by its nature instrument-dependent, showing
mostly features of the instrumental response, not related to the physics of the source.

\section{Spectra and spectral states of black-hole binaries}
\label{states}

\subsection{Classification of the states}
\label{classification}

The simplest classification of spectral states of black-hole binaries consists
of two states, the hard and the soft. Very roughly, the photon spectral indices
around $\sim 10$ keV during those states are $\Gamma\simless 2$ and $\simgreat
2$, respectively. The hard state (also called low, LS) is characterized by a
weak soft blackbody component and a high-energy spectral cutoff at
$\simgreat$100--200 keV, whereas the soft state is characterized by a strong
blackbody component and no observable (as yet) high-energy cutoff in the
spectral tail beyond the blackbody. The hard state can take place at a very
large range of the Eddington ratio, up to $L/\ledd\sim 0.2$--0.3\cite{dg03},
where $\ledd\simeq 1.5\times 10^{38} (M/\msun)$ erg s$^{-1}$. Weak (`off')
states have usually relatively hard spectra as long as they can be measured, so
they probably are a subset of the hard state. However, the detailed form of
spectra of weak states remains uncertain, with some observations showing
$\Gamma\simgreat 2$, e.g., in XTE J1118+480\cite{mcclintock03} or GX
339--4\cite{z04}.

Then, the soft state is often subdivided into the high (HS), very high (VHS) and 
intermediate (IS) states. The `pure' high, or ultrasoft, state is supposed to 
have only the blackbody disk component, without a high-energy tail, but this 
definition is clearly dependent on the sensitivity of the instrument used. In 
fact, most of spectra claimed to represent this state have high-energy tails at 
some level, e.g., all of those presented in Ref.~\citen{mr04}. The IS takes 
place during transitions between the hard and soft states, and spectrally it is 
characterized by the high-energy tail starting close to the peak of the disk 
blackbody. Then, the VHS is often defined as having the same spectral type as 
the IS\cite{gd03}, but at high luminosities. In fact, the VHS usually takes 
place during transitions from the hard state to the soft state, so the 
difference with respect to the IS is small, if any\cite{r99,h01}. The VHS/IS can 
take place at a large range of the luminosity, between $\sim 0.02\ledd$ or even 
less, e.g., during state transitions of Cyg X-1, and $\simgreat 0.3\ledd$ (e.g., 
in XTE J1550--564, GX 339--4, GRS 1915+105). Then, the HS (with a weak tail) can 
also take place at any luminosity $\simgreat 0.02\ledd$\cite{m03} up to $\sim 
\ledd$, e.g., in the $\gamma$ state (in the classification of Ref.~\citen{b00} 
of GRS 1915+105\cite{z01}.

Overall, various states can take place at various luminosities, being a
function of both the accretion rate, $\dot M$, in Eddington units, $\dot m$,
{\it and\/} the preceding behavior of the source (which gives rise to
hysteresis, see below). The presence of a high-energy tail in a soft state
appears universal, but it can be at virtually any level with respect to the
disk blackbody. Below, we give some examples.

\subsection{Cyg X-1}
\label{ss:cygx1}

\begin{figure}[t!]
\centerline{\includegraphics[width=10.5cm]{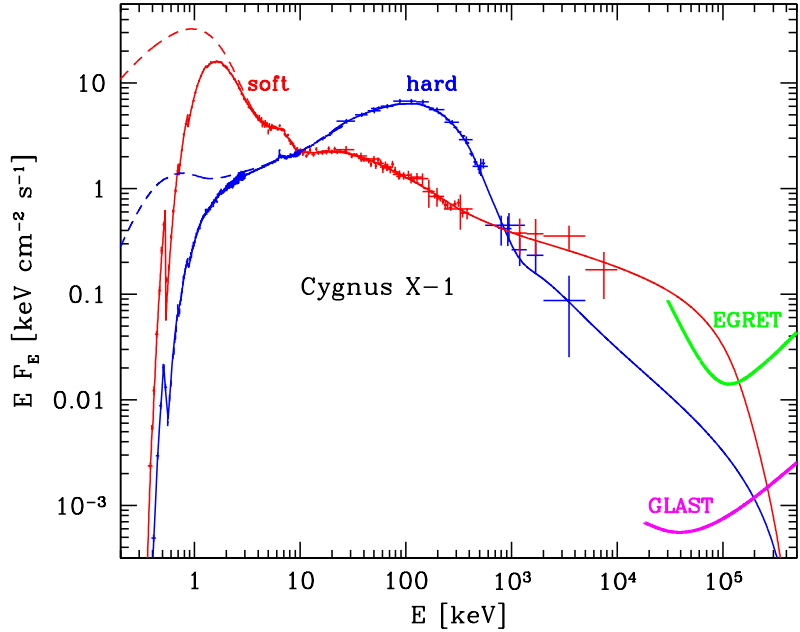}}
\caption{Spectra\cite{mc02} of the soft state of 1996 June and of the average
hard state of Cyg X-1. The models\cite{coppi99,g99} (solid curves) in both
states consist of hybrid Comptonization of disk blackbody photons, Compton
reflection and fluorescent Fe K emission. The dashed curves give the model
spectra before absorption. We also show the upper limit from the EGRET
observations in the hard state and an expected sensitivity of \glast, see \S
\ref{ss:cygx1} for details. See astro-ph for color versions of this and
subsequent figures. } \label{cygx1}
\end{figure}

\begin{figure}
\centerline{\includegraphics[width=9cm]{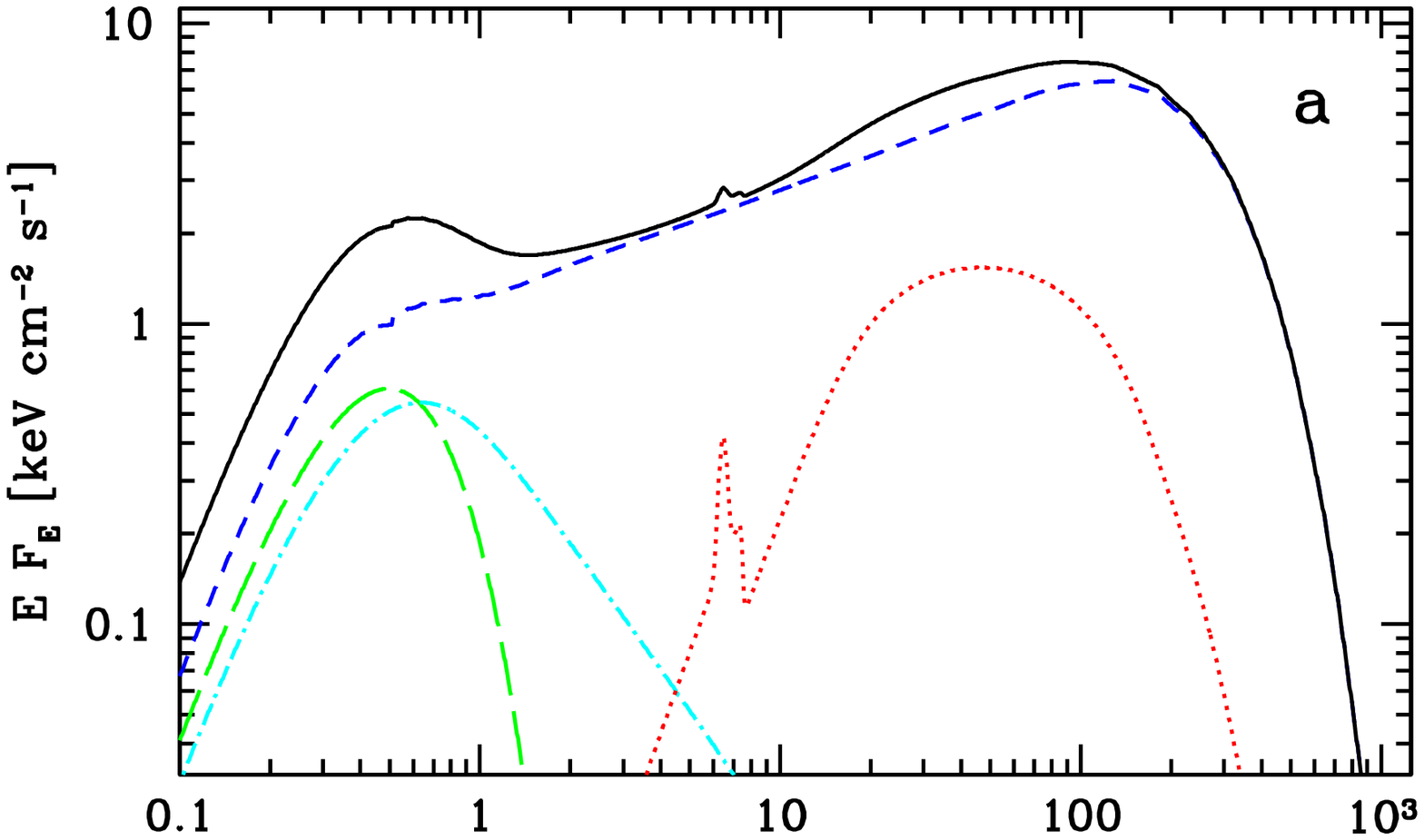}}
\centerline{\includegraphics[width=9cm]{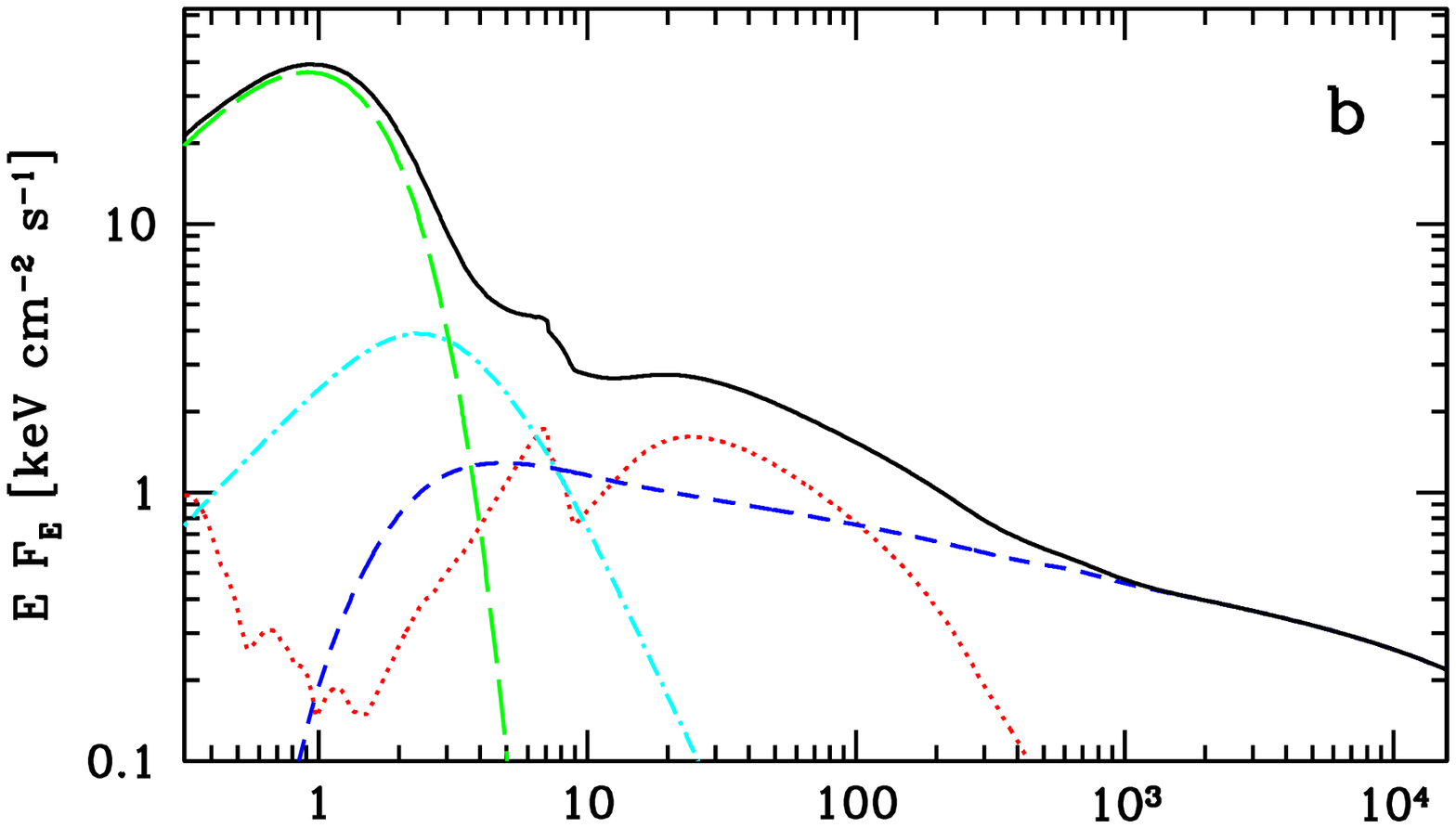}}
\centerline{\includegraphics[width=9cm]{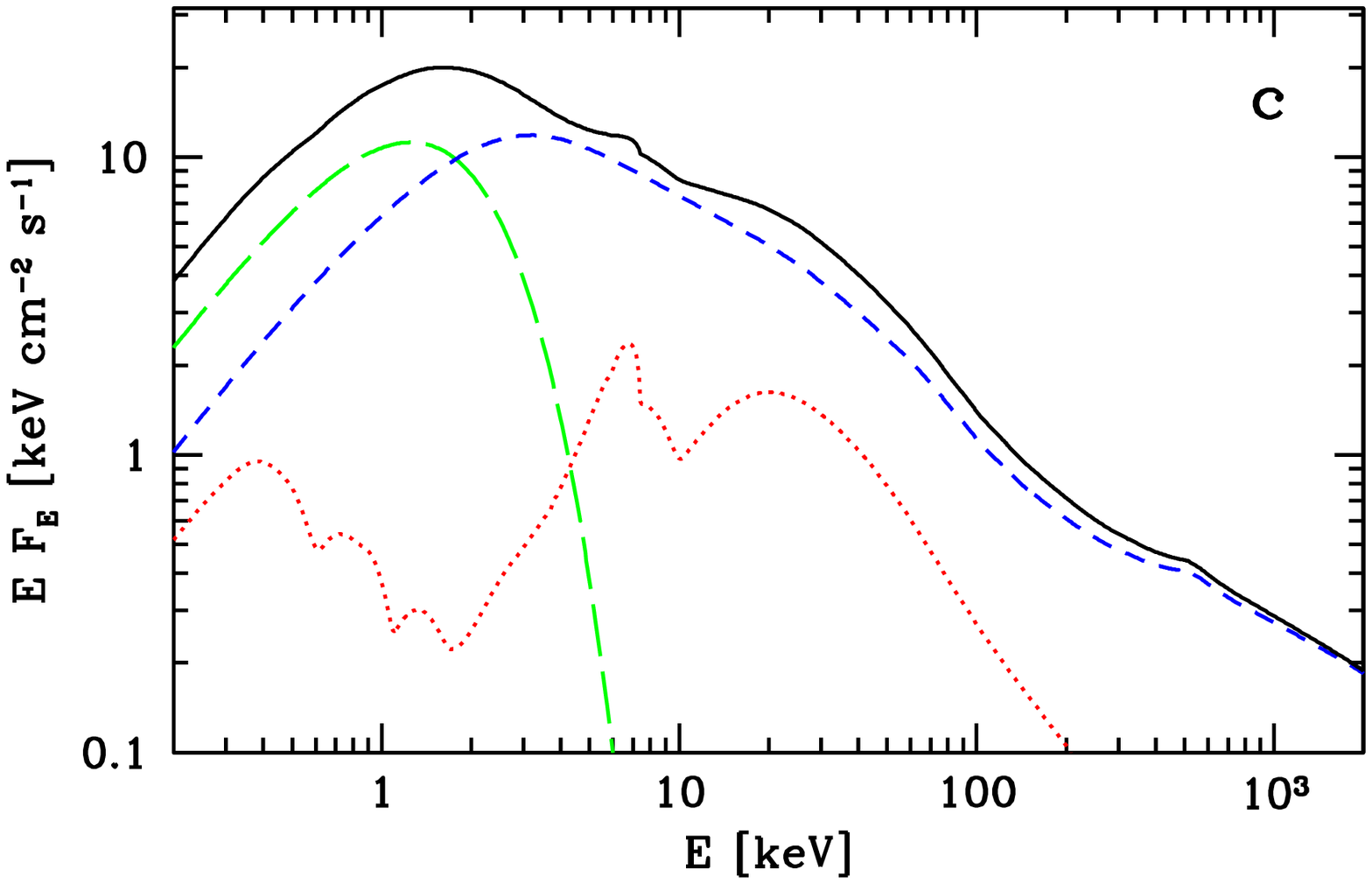}}
\caption{Components of fits to typical (a) hard and (b) soft state spectra of Cyg X-1, and (c) a VHS/IS spectrum of XTE J1550--564\cite{gd03}. The hard-state spectrum is from an observation by \sax\cite{d01}, and the soft-state is from a multi-instrument observation in 1996 June\cite{mc02}. All the spectra are intrinsic, i.e., corrected for absorption, and solid curves give the total spectra. The (green) long dashes, and (red) dots correspond to the unscattered blackbody and Compton reflection/Fe K$\alpha$ fluorescence, respectively. The (blue) short dashes give the main Comptonization component, which is due to scattering by (a) hot thermal electrons with $kT\simeq 75$ keV and $\tau\simeq 1$, (b) nonthermal electrons of a hybrid distribution, and (c) the total hybrid distribution. The (cyan) dot-dashes show (a) the soft excess, seen in the hard state of Cyg X-1 and some other sources, most likely due to Compton scattering in a region spatially different from that of the main hot plasma, and (b) scattering by thermal electrons of the hybrid distribution.
}
\label{cygx1_comp}
\end{figure}

Figure \ref{cygx1} shows the two main states of Cyg X-1\cite{mc02}. The spectrum
of the hard state peaks at $\sim 100$--200 keV, which property is very
characteristic of this state in black-hole binaries\cite{g97,g98,zpm98,w02}.
There is by now a very strong body of evidence that the dominant radiative
process in this state is Comptonization of disk blackbody photons (at the
characteristic temperature of a fraction of a keV) by thermal electrons in a
plasma at a temperature of $kT_{\rm e}\sim 50$--100 keV and a Thomson optical
depth of $\tau \sim 1$\cite{g97,zpm98,d01,f01a,f01b,w02,z02,z03}. The power in
hard photons from the Comptonizing plasma, $L_{\rm H}$, is $\gg L_{\rm S}$, the
power in the soft blackbody photons. A brief history of the thermal Compton
model applied to accreting black holes is given, e.g., in Ref.~\citen{z00}.  In
addition, there is a secondary, reprocessing, component due to Compton
reflection from a cold medium\cite{mz95} accompanied by Fe K fluorescent
emission\cite{g97,zpm98,d01,f01a,w02}. The reflector is likely to be the same
accretion disk that gives rise to the disk blackbody emission. The reprocessing
component is sometimes relativistically broadened\cite{d01}, especially, in
softer of the hard states of Cyg X-1\cite{f01a,m02}. The components of a
hard-state spectrum of Cyg X-1 are shown in Figure \ref{cygx1_comp}a.

Although the scattering plasma is predominantly thermal in the hard state, there
are indications that the electron distribution has some high-energy tail, i.e.,
it is hybrid, thermal/nonthermal. The main evidence for it has been provided by
\gro/COMPTEL observations of Cyg X-1 in the hard state\cite{mc00,mc02}. The
photon tail beyond the thermal-Compton spectrum is relatively weak, as shown
in Figure \ref{cygx1}.

The spectra in the soft state appear to contain the same ingredients as those in
the hard state, but there are different proportions between the disk blackbody
and Comptonization\cite{pkr97} and between the thermal and nonthermal parts of
the electron distribution in the Comptonizing plasma\cite{pc98}. Opposite to the
hard state, the disk blackbody emission is energetically dominant, $L_{\rm S}\gg
L_{\rm H}$, and the nonthermal electron tail is much
stronger\cite{g99,z00,z01,z02,gd03}. These properties are illustrated by the
soft-state spectrum of Cyg X-1 shown in Figure \ref{cygx1}, with the components
of the fit shown in Figure \ref{cygx1_comp}(b). We see that Comptonization by
nonthermal electrons dominates at $E\simgreat 10$ keV. The Compton reflection
and Fe K fluorescence are present as well, and their strength, the ionization of
the reflector, and the degree of relativistic broadening are all larger than
those in the hard state\cite{g99,gcr99}.

No high-energy cutoff in the power-law tail of the soft state has been detected
as yet\cite{g99,z01,mc02}. The highest observed energy of the nonthermal photon
power law is $\sim$10 MeV, measured in Cyg X-1 by the \gro/COMPTEL\cite{mc02},
see Figure \ref{cygx1}.

\begin{figure}[b!]
\centerline{\includegraphics[width=11cm]{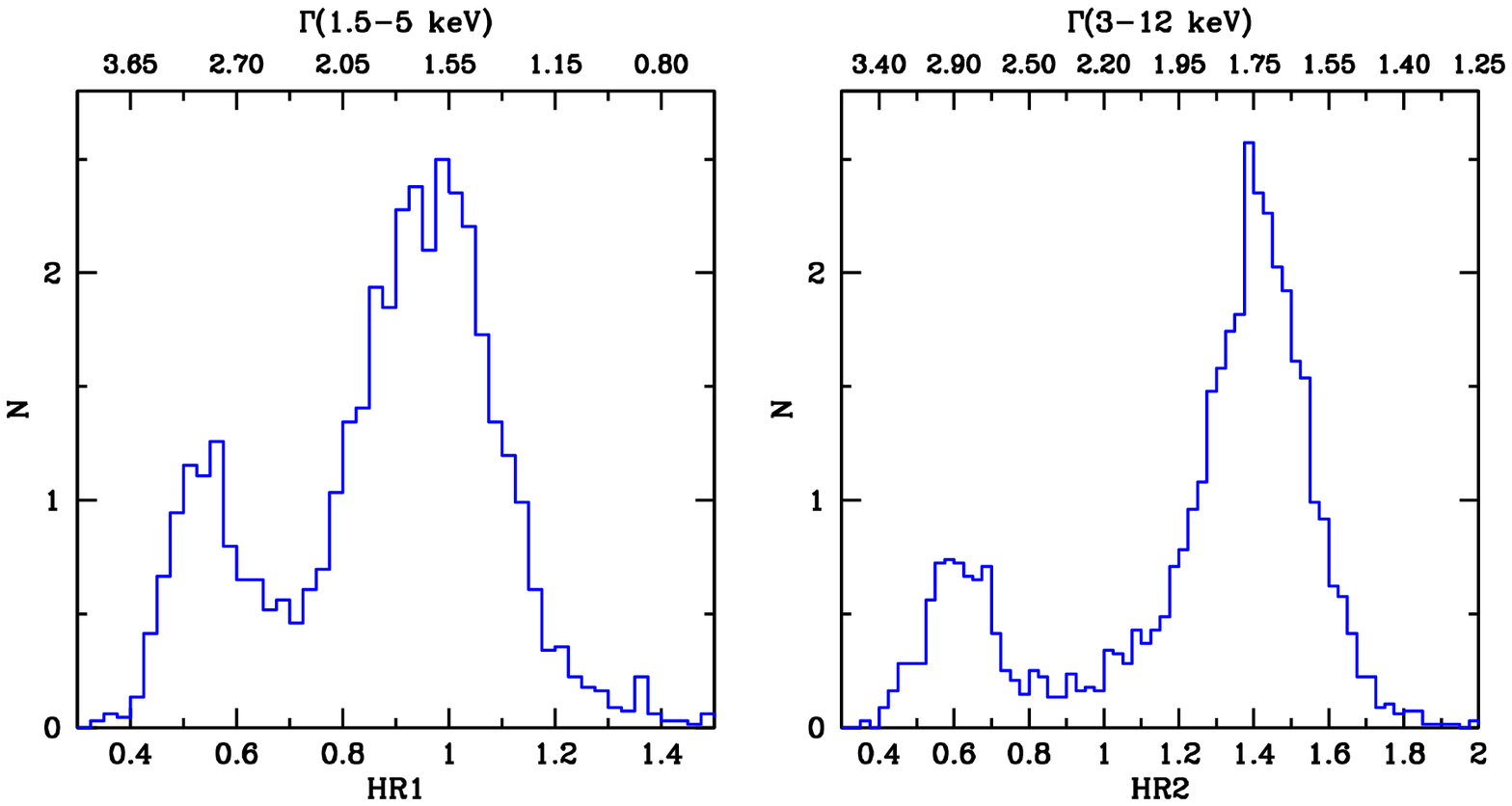}} \caption{Histograms
(normalized to unity) of the distributions of the \xte/ASM hardness ratios in
the bands of 1.5--3--5--12 keV observed from Cyg X-1. The upper axes show the
corresponding values of the spectral index of a power law going through the two
adjacent energy bands and yielding the observed hardness ratio\cite{z02}
(without correcting for absorption). We clearly see bimodal distributions
corresponding to the hard and soft states. } \label{cygx1_hist}
\vskip 0.5cm
 \centerline{\includegraphics[height=7.6cm]{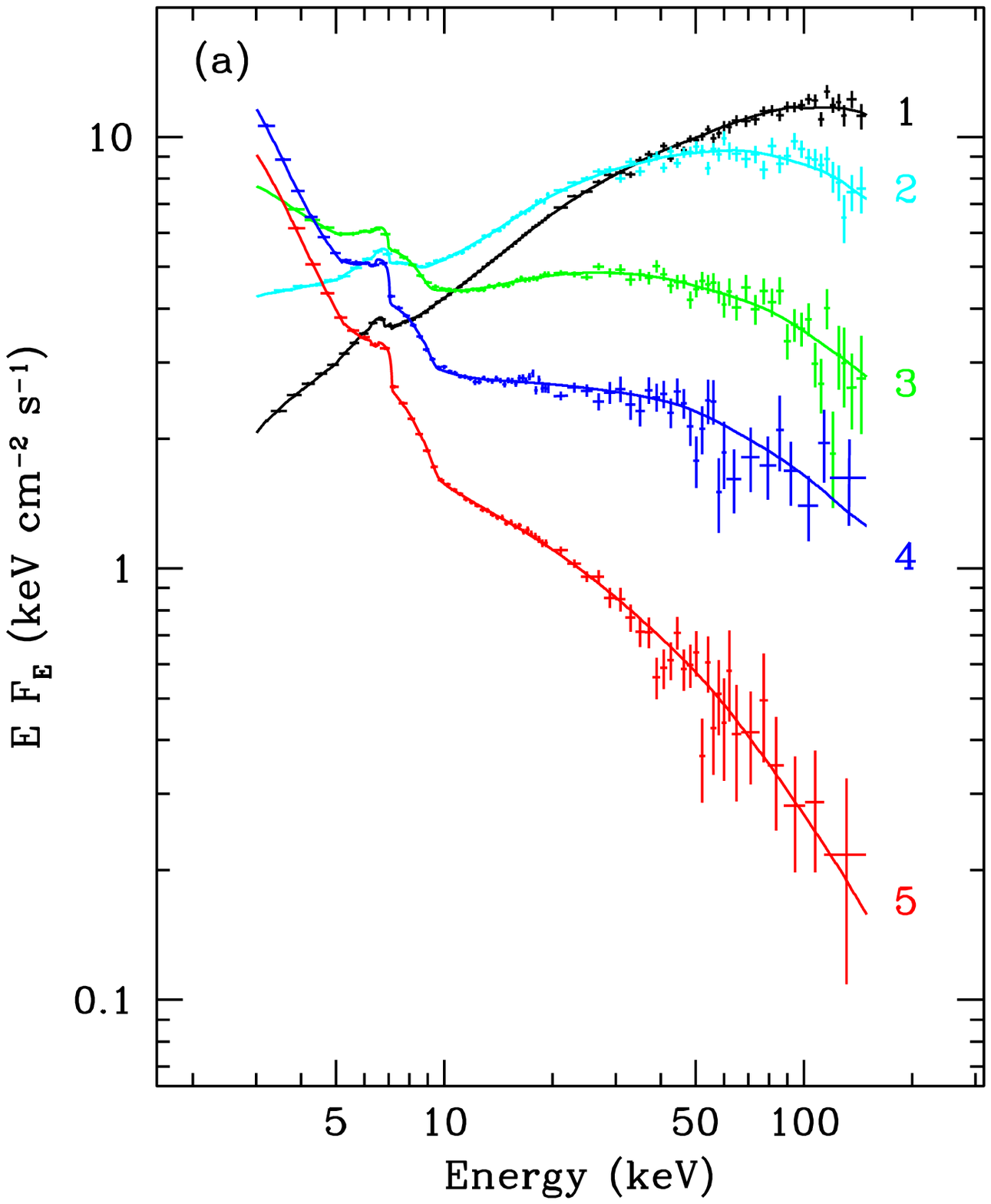}\hfill
\includegraphics[height=7.6cm]{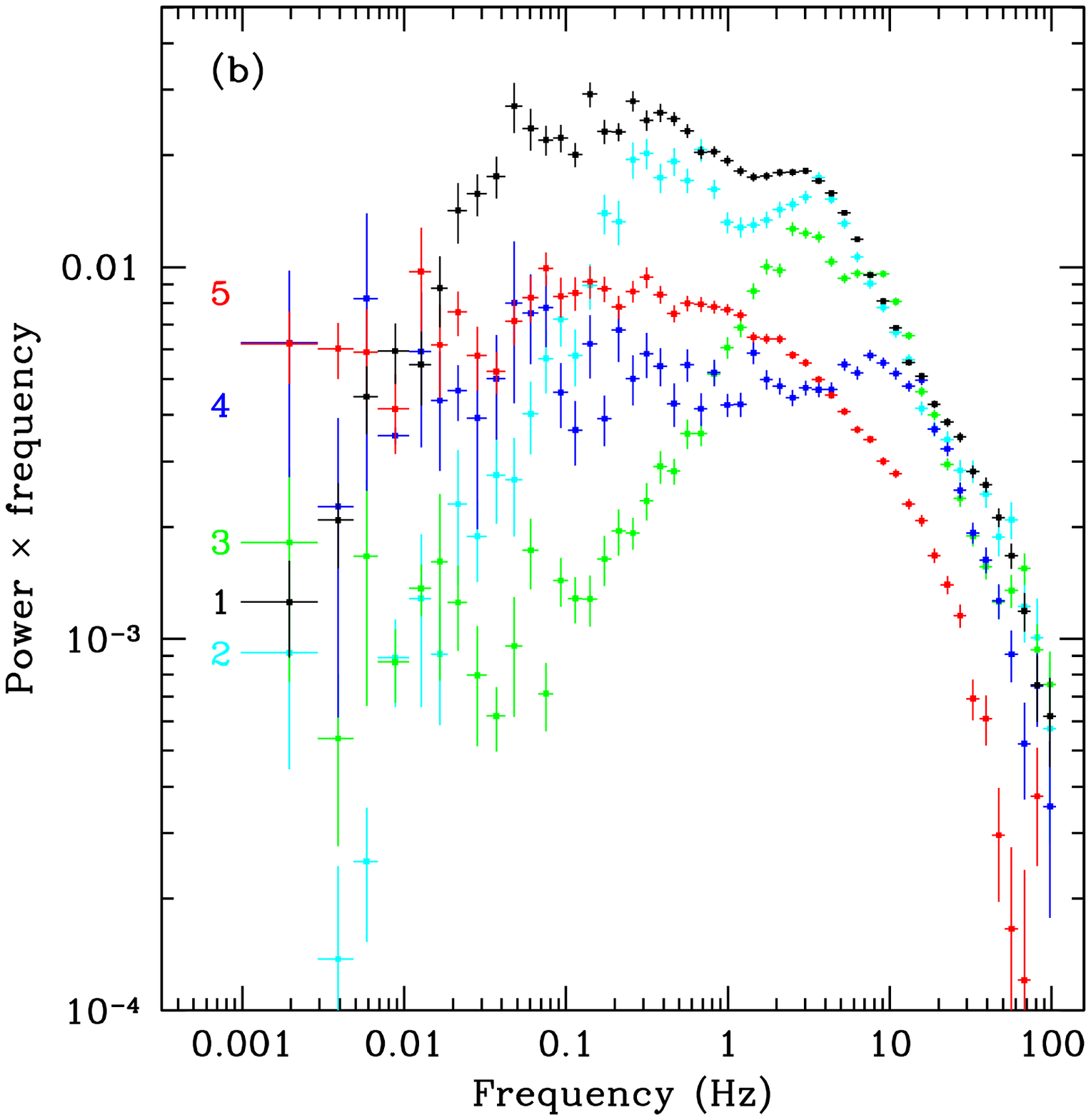}} \caption{The range of energy
(a) and power (b) spectra of Cyg X-1 measured by pointed \xte\/ observations.
The models of the PCA/HEXTE data shown in (a) involve hybrid Comptonization of
disk blackbody emission and reflection, see \S \ref{ss:cygx1} for details. The
power spectra correspond to the $\sim$2--60 keV range and have been corrected
for dead time of the PCA. } \label{cygx1_sp}
\end{figure}

It is of major importance where the high-energy cutoff in the soft state is and
how it can be measured. Figure \ref{cygx1} shows a sensitivity of
\glast\cite{michelson03} for a $4\times 10^5$ s net exposure\footnote{For the
effective area of
www-glast.slac.stanford.edu/software/IS/glast\_lat\_performance.htm and
requiring 20 photons per unit $\ln E$. This relatively large number is intended
to approximately take into account the presence of the strong diffuse emission
in the Galactic plane region.}. We see that \glast\/ will be able to strongly
constrain the high-energy spectra of Cyg X-1 in both states. The model
spectra\cite{mc02} shown here assume the maximum Lorentz factor of the power-law
electrons (beyond the respective Maxwellian distributions) in each state of
$10^3$. If this Lorentz factor were higher or lower, the cutoffs would be at
somewhat higher or lower energies, respectively, than the ones shown. On the
other hand, the models use the compactness parameter, $\ell$ ($\propto L/R$,
where $R$ is the characteristic size of the Comptonizing plasma, see, e.g.,
Ref.~\citen{g99} for details of the definition), fitted in Ref.~\citen{mc02} to
have relatively low values of $\sim$4, $\sim$30 for the soft and hard state,
respectively. If these parameters were higher, the high-energy cutoffs would
appear at lower energies due to absorption in photon-photon production of \ee\
pairs.

The current upper limit to the $\sim$100 MeV flux from Cyg X-1 is
from the \gro/EGRET\cite{mc00,h99}, and it is also shown in Figure \ref{cygx1}.
We have taken into account the dependence of the effective area on
energy\footnote{From cossc.gsfc.nasa.gov/egret/egret\_tech.html.} and normalized
it to the position-dependent upper limit  for the number of photons $\geq$100
MeV\cite{h99}. Unfortunately, the EGRET performed rather few observations of Cyg
X-1, and, in particular, it did not observe Cyg X-1 during its 1996 bright soft
state (shown in Fig.\ \ref{cygx1}).

Given the sensitivity of \glast, we expect to be able to put strong constraints
on the presence of hadronic processes in Cyg X-1. If protons are accelerated to
high energies together with electrons, proton collisions will produce pions.
Then, decay of neutral pions gives rise to a spectral hump around $\sim$100 MeV
(see, e.g., the models of figures 2--4 of Ref.~\citen{z86}. Also, decay of
charged pions produces additional relativistic pairs, which further modify the
\g-ray spectrum. This issue was studied by Ref.~\citen{b03}, who found that the
measurements\cite{z01} of GRS 1915+105 at high energies constrain the fractional
power in nonthermal protons to $\simless 5\%$. Still, though hadronic processes
may not play the dominant role, measuring their contribution is of major
importance for our understanding of the physics of black-hole binaries.

\begin{wrapfigure}{1}{6.6cm}
\centerline{\includegraphics[width=7.cm]{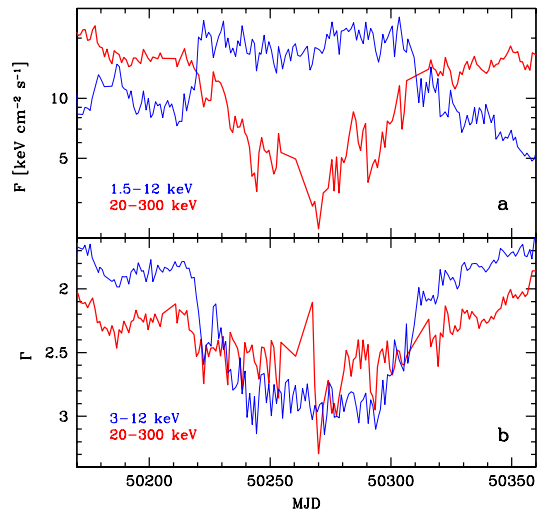}\hfill} \caption{(a)
Lightcurves\cite{z02} from the ASM and BATSE of the 1996 state transitions of
Cyg X-1, showing no substantial hysteresis. (b) The average spectral
indices\cite{z02} in two bands. } \label{cygx1_96}
\end{wrapfigure}

The distribution of the states of Cyg X-1 is bimodal, see Figure
\ref{cygx1_hist}. Figure \ref{cygx1_sp}(a) shows the range of spectra (observed
over the timescale of hours) observed so far from Cyg X-1 by the PCA and HEXTE
on board of \xte. The spectra 1--2 belong to the hard state, and we see the
characteristic pivot at $\sim$30 keV within this state\cite{z02}. The spectrum 4
is from the 1996 soft state, and it is very similar to that shown in Figure
\ref{cygx1}. On the other hand, spectra observed during the long 2002 soft state
were often much softer\cite{gz03}, occasionally as soft as the spectrum 5. Then,
the spectrum 3 is an intermediate one, measured during the 1996 hard-to-soft state transition\cite{g99}.

Figure \ref{cygx1_sp}(b) shows the power spectra corresponding to those energy
spectra. We see the hard-state spectra (1--2) have the characteristic broad
maximum in the $\sim$0.1--3 Hz in the power per logarithm of frequency,
sometimes with two peaks, see e.g., Ref.~\citen{gcr99}. The upper peak is
usually referred to as the low-frequency QPO, while the lower one, as the break
frequency. The soft states (4--5) is characterized by a flat (in power per
$\log f$) power law with a high-frequency cutoff above $\sim$1--10 Hz, see,
e.g., Ref.~\citen{cgr01}. On the other hand, the intermediate state (3) shows a
relatively narrow single peak (a QPO) at $\sim$3 Hz, often seen in IS/VHS power
spectra\cite{h01}.

The transitions to and out of the soft state in Cyg X-1 take place at about the
same flux level, as illustrated in Figure \ref{cygx1_96}. The hard state is
observed at $L\simeq (0.01$--$0.02)\ledd (d/2\,{\rm kpc})^2 (10\msun/M)$, and
higher luminosities always correspond to intermediate and soft states\cite{z02}.
The luminosity of the soft-state spectrum shown in Figures \ref{cygx1},
\ref{cygx1_comp}b is $L\simeq 0.05\ledd (d/2\,{\rm kpc})^2
(10\msun/M)$\cite{z02}.

\subsection{Hysteresis in Low-Mass X-ray Binaries}
\label{hysteresis}

Unlike the case of Cyg X-1, where the transitions between the soft and hard
states take place approximately at the same $L$ (which corresponds to a
one-to-one correspondence between the luminosity and the spectral state), state
transitions in LMXBs can take place in a wide range of $L/\ledd$, and,
consequently, various spectral states can correspond to the same
$L/\ledd$\cite{m95,mc03,r04,z04}. This is obviously related to the existence of
ranges of $L/\ledd$ and radii at which two stable accretion solutions,
optically thick (soft)\cite{ss73} and optically thin (hard)\cite{sle76,ny95},
exist.  The maximum possible luminosity of the hard state (probably limited by
cooling of the hot plasma) is larger than the minimum possible luminosity of
the soft state (probably limited by evaporation of the
disk\cite{meyer00,rc00}). The state of a variable source in a given moment is
then determined not solely by $L/\ledd$ but also by its history. This leads to
a hysteresis (lagging of the effect in a physical system when the acting force
is changed) in the lightcurve of the source; the hard-to-soft state
transitions during the rise phase occur at higher luminosities than the
soft-to-hard one during the decline phase.

\begin{figure}[b!]
\centerline{\includegraphics[height=9.3cm]{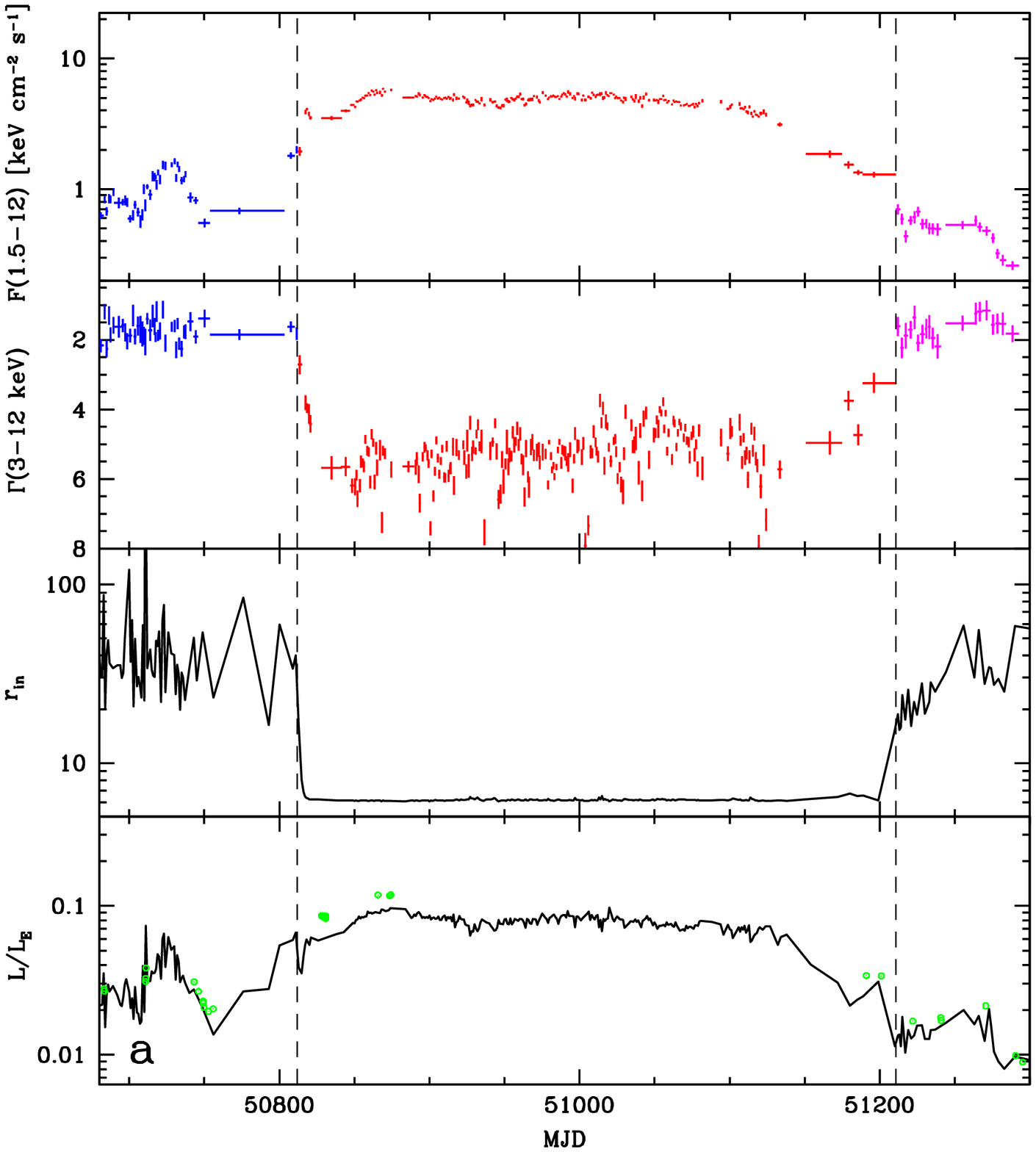}
\includegraphics[height=9.3cm]{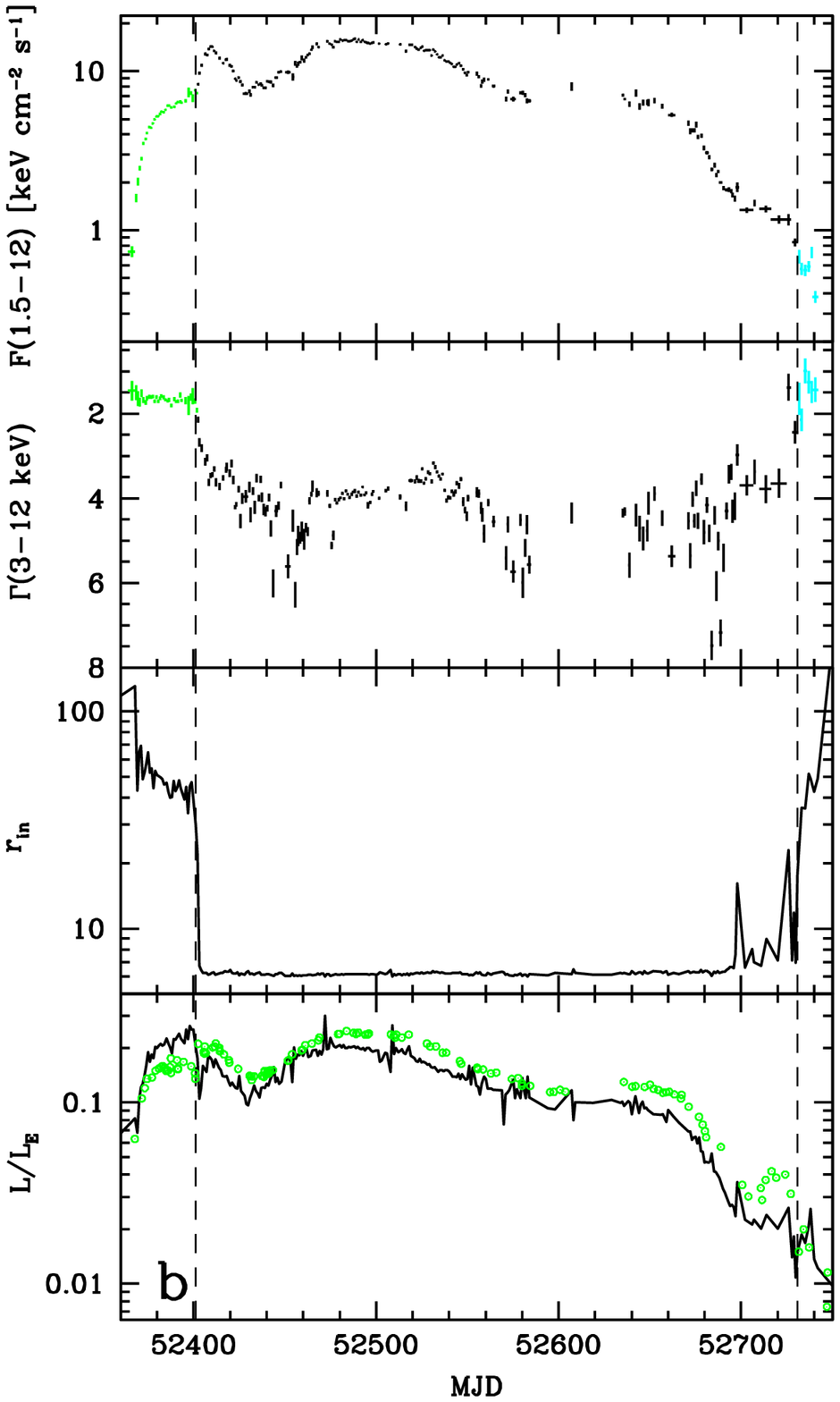}}
\caption{The 1.5--12 keV \xte/ASM flux and the 3--12-keV photon spectral index
with the values of $L/L_{\rm E}$ and the inner radius of the accretion
disk, $r_{\rm in}$ fitted by a simple accretion model (for $M=10\msun$ and $d=8$ kpc) as a function of time during the two main outbursts of GX
339--4\cite{z04}. State transitions are marked by the vertical lines. The circles in the bottom panels show $L/L_{\rm E}$ based on the available PCA/HEXTE data\cite{z04}, which are in a relatively good agreement with the ASM model. We see that whereas the transitions out of the soft state take place at $\sim 0.02\ledd$, the transitions from the hard to the soft state take place at substantially higher $L$, which are also different for each of the two outbursts. }
\label{gx339_hysteresis}
\end{figure}

The presence of pronounced hysteresis in LMXBs is illustrated in Figure
\ref{gx339_hysteresis}, showing the history of the two recent outbursts of GX
339--4\cite{z04}. The upper panels show the ASM lightcurve and the 3--12 keV
spectral index (which clearly identifies the spectral state in a given moment).
The lower panels show the corresponding inner radius of the optically-thick disk
and the bolometric $L$, the latter estimated from an accretion model and
available PCA/HEXTE data\cite{z04}. We see that the transitions from the
hard to the soft state took place at much higher fluxes and luminosities than
the transitions from the soft to the hard state, in agreement with our
discussion above. Both soft-to-hard transitions took place at about the
same $L\sim 0.02\ledd$, also typical for other black-hole
binaries\cite{m03}. Thus, the history determining the moment of the transition
appears to be simply the pre-existence of an optically-thick inner disk,
presumably extending close to the minimum stable orbit.

However, the transition out of the hard state during the second outburst took
place at $L\sim 0.2\ledd$, a few times higher than $L\sim 0.07\ledd$ for the
first outburst. This is also in agreement with data for other
LMXBs\cite{mc03,r04}, showing this type of transition taking place at a wide
range of $L/\ledd$. The reason for the variety of the transition $L/\ledd$ has
to be related to the structure of the hot accretion flow being more complex than
that of the cold disk extending to the minimum stable orbit. This additional
complexity is likely to be the value of the inner radius of an outer cold disk,
most likely existing in this type of flow. In the example of Figure
\ref{gx339_hysteresis}, the first transition followed a $\sim 10^3$ d of
sustained hard-state activity of GX 339--4\cite{z04}. The cold disk extending
relatively close to the black hole, $r_{\rm in}\equiv R_{\rm in}c^2/GM\sim
10^2$, was then probably built by viscous processes long time before the
transition towards the soft state started. Then, it took a relatively short time
for this cold disk to build all the way down to the minimum stable orbit after
$\dot M$ started to increase. On the other hand, the second outburst took place
after a $\sim 10^3$ d of quiescence. That outburst presumably started in outer
regions ($r_{\rm in}\gg 10^2$) with the standard H ionization instability. The
buildup of the inner cold disk took longer than in the previous case, which
allowed the hot flow to exist up to a later stage of the increasing-$\dot M$
phase, i.e., up to a higher $L$.

As shown in Figure \ref{cygx1_96}, there is no noticeable hysteresis in Cyg X-1.
This is clearly related to this object being persistent and with a high-mass
companion. This may be directly due to its relatively narrow range of observed
$L$, in which case the outer cold accretion disk in the hard state presumably
stays relatively close to the black hole. It may be also related to the outer
disk radius being relatively small due to the accretion process being
quasi-spherical\cite{bi01,ib01}.

Apart from the uncertain details of the accretion processes, we have a firm
observational fact that transitions between the hard and the soft state can
happen in a wide range of $L/\ledd\sim 0.01$--0.2. Correspondingly, the
IS/VHS can also take place at various values of $L/\ledd$.

\subsection{The nature of the high and very high states}
\label{ss:vhs}

\begin{figure}[b!]
\centerline{\includegraphics[height=6.cm]{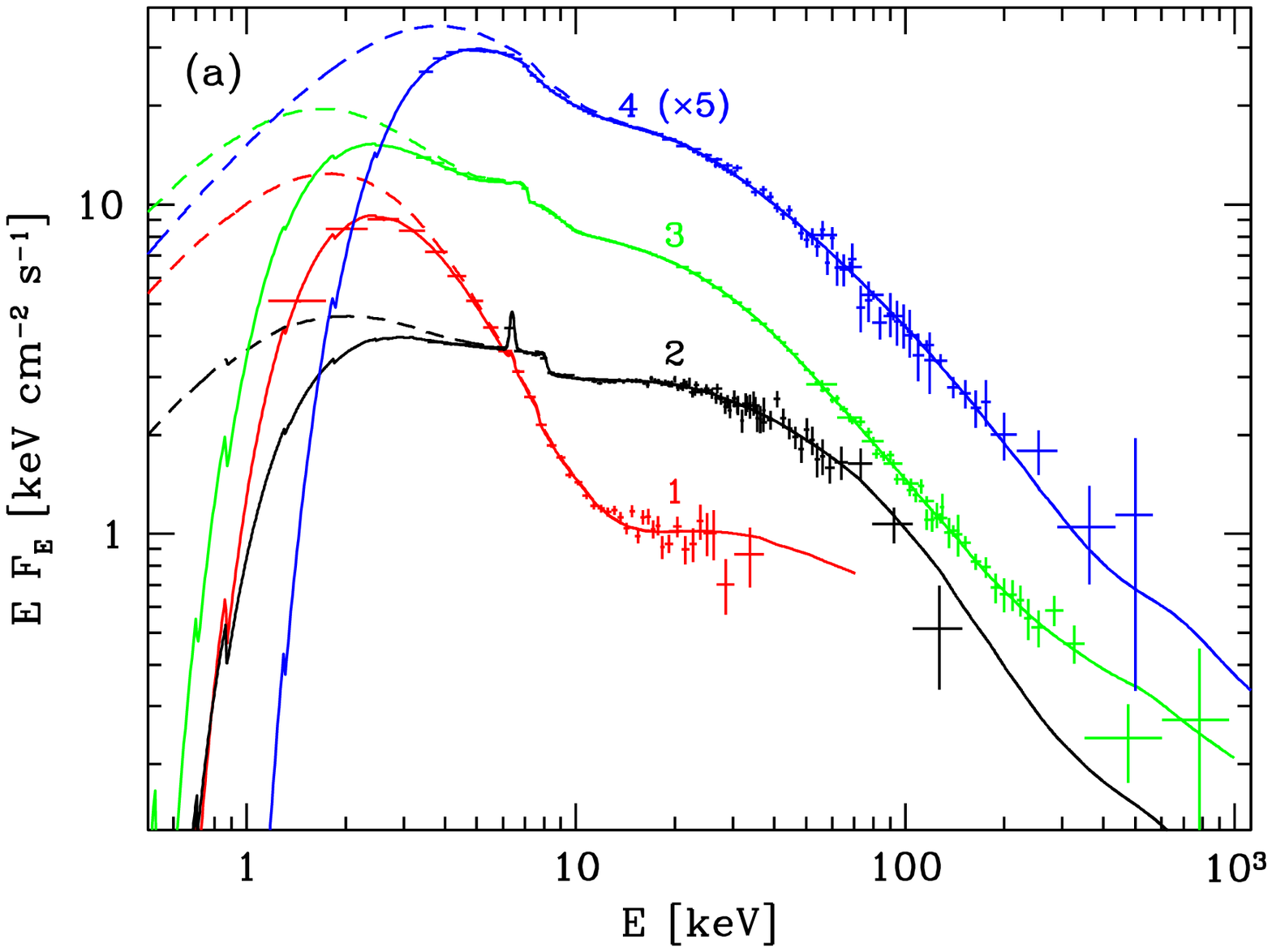}\hfill
\includegraphics[height=6.cm]{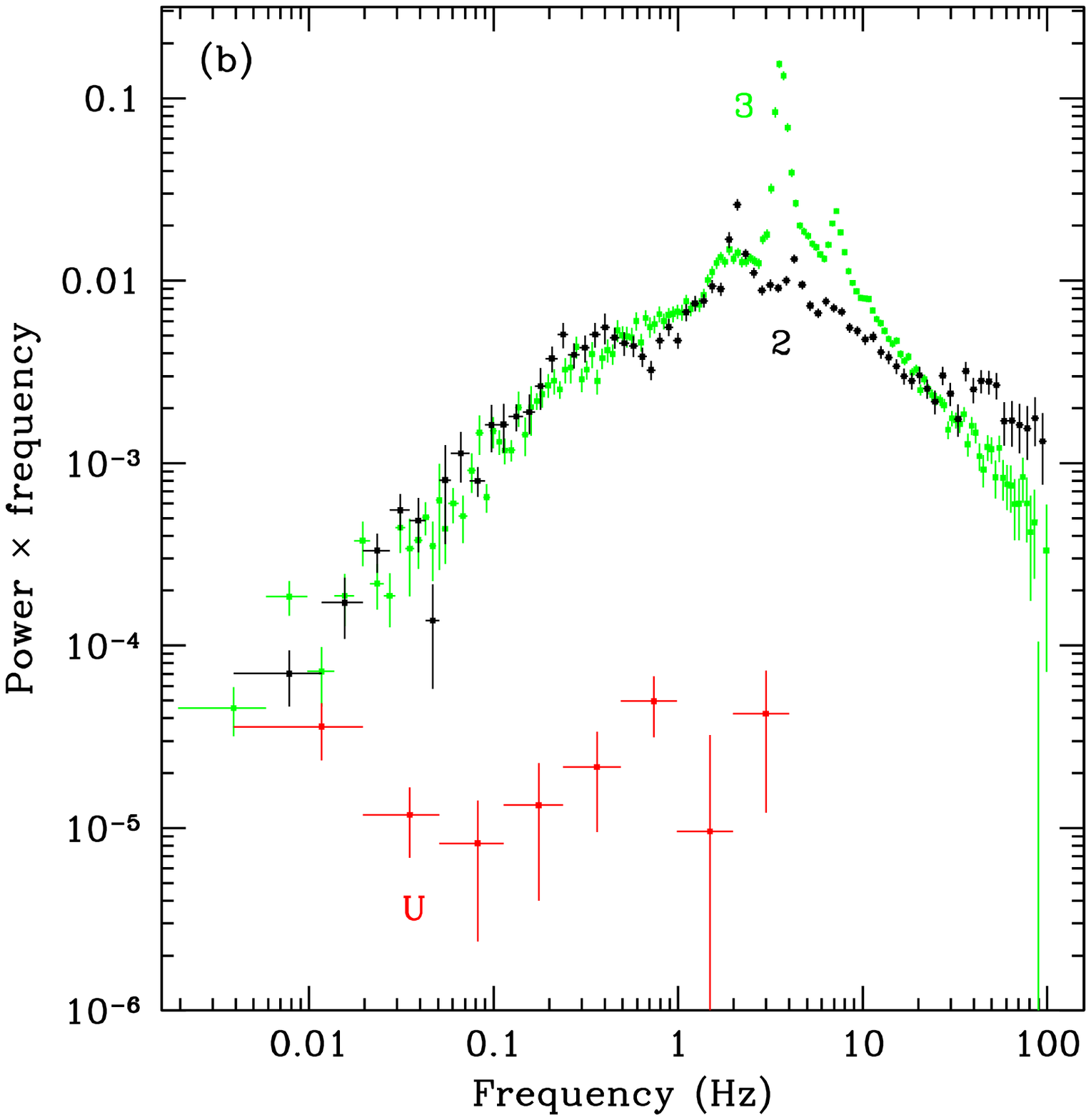}}
\caption{(a) Various spectra classified in astrophysical literature to
represent the VHS. The dashed curves represent the intrinsic spectra before
absorption. The spectrum 1 is from a \ginga\/ observation of GX 339--4 on MJD
47772 and belongs to the original spectral state classified as representing the
VHS\cite{m91}. The spectrum 2 is from an observation of GX 339--4 on MJD 52402,
classified to belong to the VHS by Ref.~\citen{s02}. The spectrum 3 is from
observations of XTE J1550--564 on MJD 51082--51091\cite{gd03}, see Figure
\ref{cygx1_comp}c for the components of the fit. The spectrum 4 (rescaled by a
factor of 5 for clarity of display) is from an observation of GRS 1915+105 on
MJD 50582--50588\cite{z01}. (b) Power spectra in the VHS states of  GX 339--4
(spectrum 2, MJD 52402) and XTE J1550--564 (spectrum 3, MJD 51086). For
comparison, the power spectrum of GX 339--4 in a high/ultrasoft state is also
shown (spectrum U, MJD 52483), see Fig.\ \ref{hs} below for its $E F_E$
spectrum. The power spectra correspond to the $\sim$2--60 keV range and have
been corrected for dead time of the  PCA. } \label{vhs}
\end{figure}

Here, we begin with a historical note on the use of the VHS term. It was
introduced by Ref.~\citen{m91} to describe a bright state of GX 339--4 observed
by \ginga. Those authors defined it by three criteria: (i) the X-ray flux
higher than that of the HS; (ii) the energy spectrum consisting of a blackbody
with a power law with $\Gamma\sim 2.5$, which they considered softer than the
high-energy tail of the soft state; (iii) rapid X-ray variability, much higher
than that in the soft state. The spectrum 1 in Figure \ref{vhs}a shows a
characteristic spectrum of Ref.~\citen{m91} (we note that those authors showed
their spectra only in count space of the \ginga\/ LAD, which made comparison
with spectra from other instruments difficult).

\begin{wrapfigure}{1}{6.6cm}
\centerline{\includegraphics[width=7.cm]{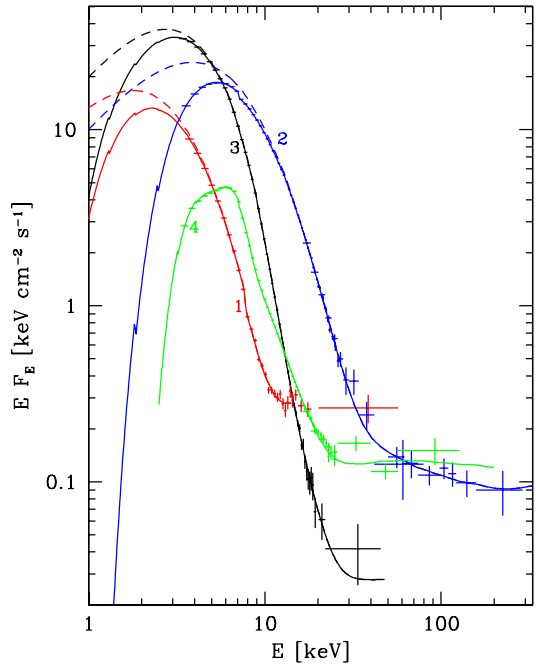}} \caption{Ultrasoft spectra
from black-hole binaries. The spectrum 1 corresponds to the brightest state of
GX 339--4 observed by \xte\/ (on MJD 52483)\cite{z04}, and its power spectrum
is shown in Fig.\ \ref{vhs}b. The spectrum 2 is from one of the brightest
states of GRS 1915+105 (on MJD 51291)\cite{z01}. The spectrum 3 is from XTE
J1550--564. The spectrum 4 is an average ultrasoft spectrum of Cyg
X-3\cite{sz04}. } \label{hs}
\end{wrapfigure}

Since that time, GX 339--4 has shown HS (ultrasoft) spectra at similar or higher 
flux level than that of the original VHS (see, e.g., the spectrum 1 in Figure 
\ref{hs}, and thus the criterion (i) cannot be used any more. We also see in 
Figure \ref{vhs} that this spectrum is very similar to the HS spectrum of Cyg 
X-1 shown in Figure \ref{cygx1}, which high-energy tail also has $\Gamma\sim 
2.5$, but which occurs at a much lower luminosity, $\sim 0.05\ledd$, several 
times below that of the VHS of GX 339--4\cite{z04}. The only visible difference 
is that the peak of the blackbody component (and presumably its maximum 
temperature) is at an energy about twice higher in GX 339--4 than in Cyg X-1. 
Thus, the criterion (ii) also cannot be used given our present knowledge of 
X-ray spectral states of black-hole binaries. Finally, although rapid X-ray 
variability is indeed often seen during the VHS, the spectrum 4\cite{z01} in 
Figure \ref{vhs} (classified as belonging to the VHS by Ref.~\citen{dg03}, was 
observed during one of the least variable states ($\chi$ in the classification 
of Ref.~\citen{b00} of GRS 1915+105, see also Ref.\ \citen{dwg04}. Thus, the 
criterion (iii) of rapid X-ray variability appears not universal as well.

Following the criterion (i) of Ref.\ \citen{m91}, it was proposed\cite{e97}
that the VHS takes place at the highest accretion rate, when the
optically-thick accretion disk extends all the way to the minimum stable orbit
but it is also surrounded by a powerful corona, dissipating most of the
available energy. However, it appears that the actual brightest states of
black-hole binaries are sometimes ultrasoft, not of the VHS type. This is the
case, e.g., in GRS 1915+105, where such an ultrasoft spectrum (number 2 in
Fig.\ \ref{hs}) was observed\cite{z01} at $L\simeq \ledd$, i.e., it was
$\sim$3 times brighter than the VHS one shown in Fig.\ \ref{vhs}. As mentioned
above, the brightest state observed by \xte\/ from GX 339--4 was
ultrasoft\cite{z04} (spectrum 1 in Figure \ref{hs}). Figure \ref{hs} also shows
the ultrasoft spectrum of Cyg X-3, which also corresponds to the brightest
state of that system\cite{sz04}.

Then, we note that the VHS appears to be almost always associated with a
transition between hard and soft states (in particular, this is the case for
spectra 2 and 3 in Fig.\ \ref{vhs} and spectrum 3 in Fig.\ \ref{cygx1_sp}a).
Rare exceptions to this rule\cite{h01} appear to be related to failed state
transitions. Thus, the VHS and IS are indeed basically identical, as suggested
by Refs.~\citen{r99,h01}. The state transition can occur at any $L\sim
(0.02$--$1)\ledd$, and the associated IS spectra may or may not correspond to
the highest bolometric luminosity in a given object. Hereafter, we will define
the IS by the presence of a soft ($\Gamma>2$) power law originating close to
the peak of the blackbody component, which definition also includes the VHS
(but not the original VHS of Ref.~\citen{m91}. Physically, such a spectrum
corresponds to comparable powers in the soft seed photons and the hard
Comptonized photons, $L_{\rm S}\sim L_{\rm H}$. Spectral components of a
typical VHS/IS spectrum are shown in Figure \ref{cygx1_comp}(c).

Furthermore, the IS can be a brief episode during a hard-to-soft transition
(e.g, in GX 339--4), or it can be an extended state showing substantial
luminosity evolution, as in the case of XTE J1550--564. In that case, the
transition from the hard state to the VHS/IS occurred at $\sim 0.25\ledd$ (for
$M=10\msun$ and $d=5.3$ kpc), after which the source continued to brighten in
the same state, reaching $\sim 1\ledd$. The source then faded, and showed a
transition to the HS only at $\sim 0.1\ledd$. This may be then classified as
the `true' VHS, corresponding to the brightest state of the source. However,
this is not the case for most of the states classified as `very high' in
literature. E.g., the transitions to the soft state in GX 339--4 during each of
the last two outbursts occurred at levels somewhat below the respective global
maxima.

Thus, we are led to a modification of the picture of Ref.~\citen{e97}, with the
VHS/IS occurring not as a unique function of $\dot m$ at its highest values,
but between the hard and soft states. Whether the VHS is or not the brightest
state of a given object depends on the time profile of $\dot m$, with the state
at a given $\dot m$ being a function of both the $\dot m$ and the preceding
behaviour of the source.

\begin{figure}[b!]
\centerline{\includegraphics[width=7.cm]{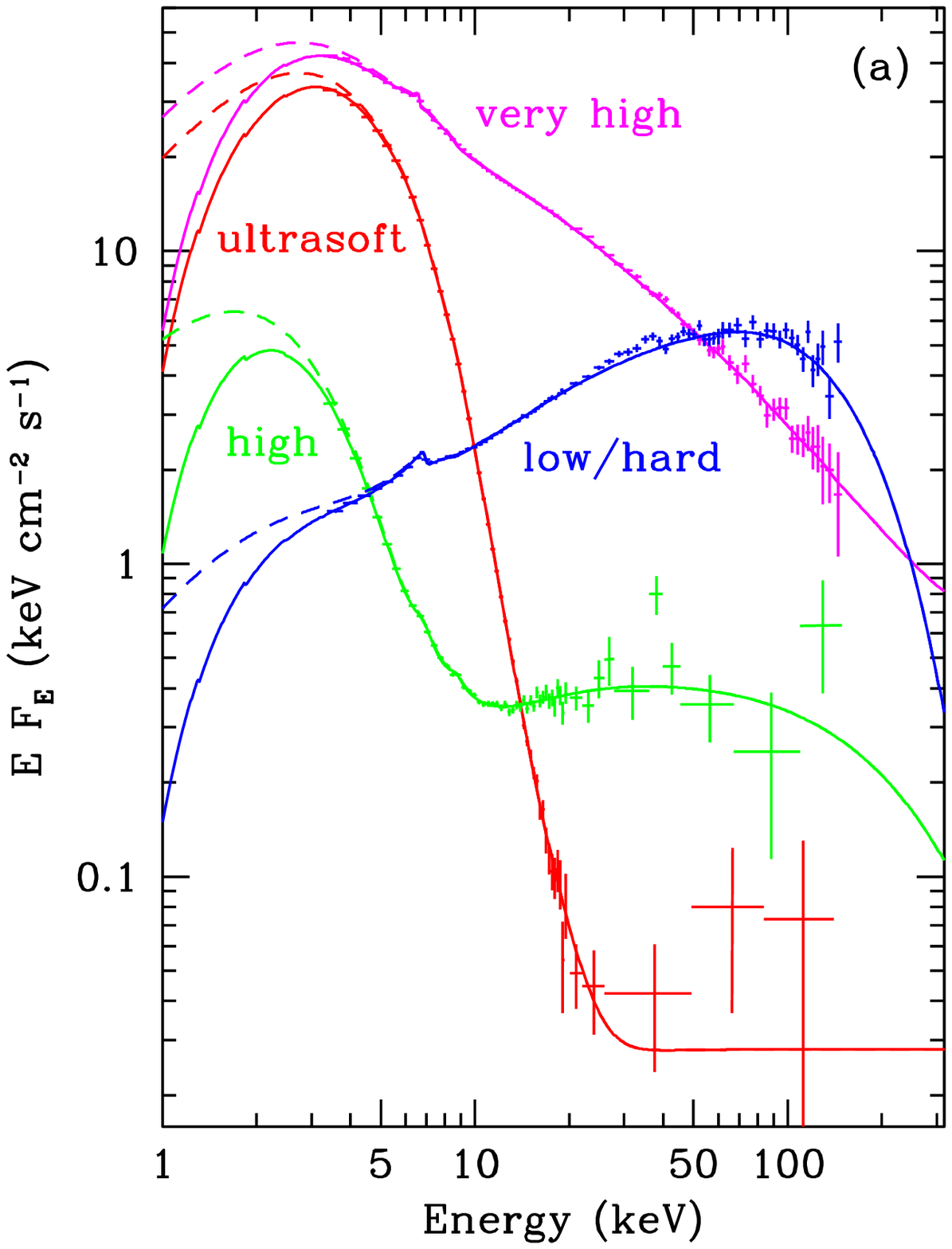}
\hfill\includegraphics[width=6.5cm]{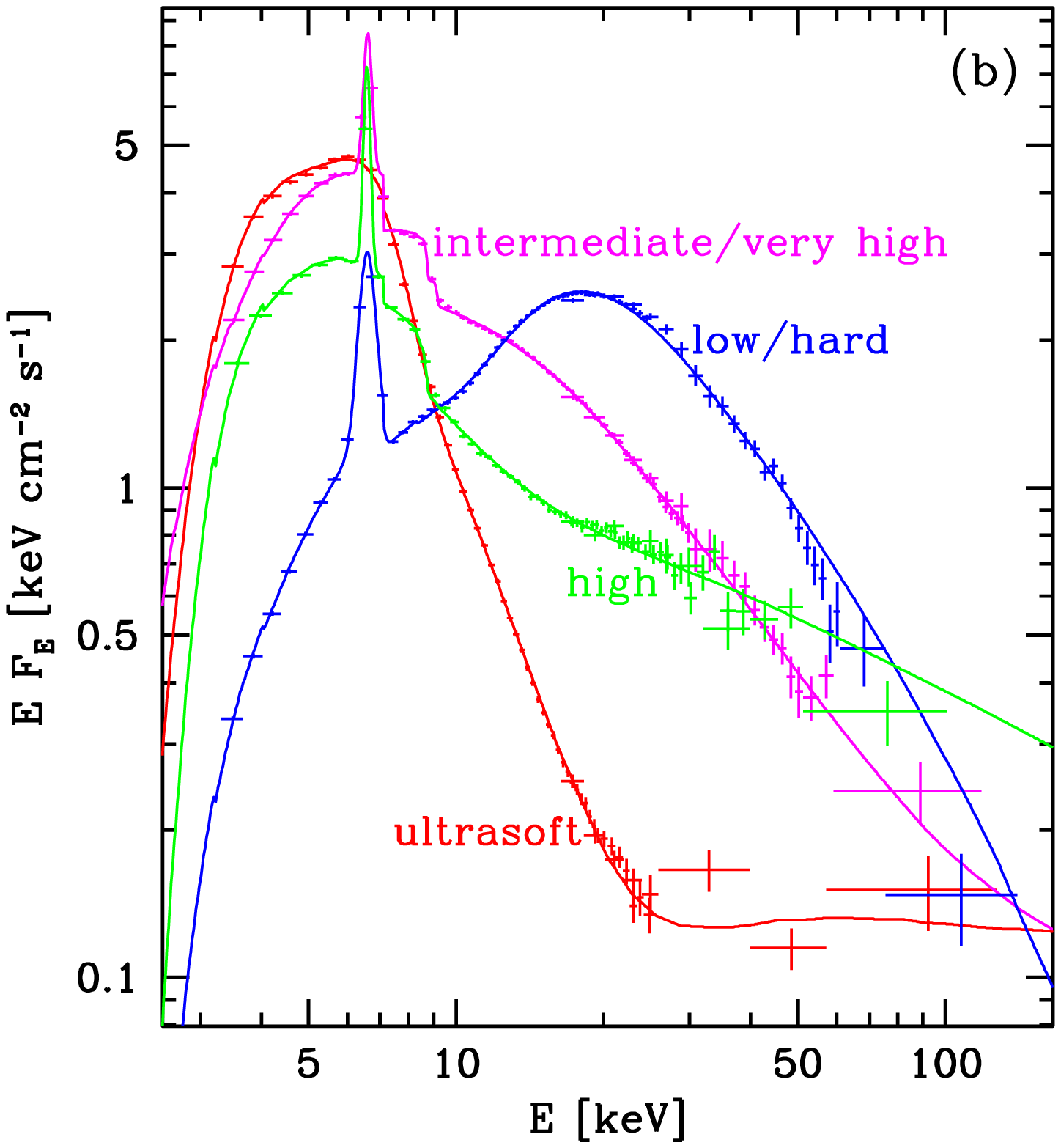}} \caption{(a)
Characteristic spectra from XTE J1550--564. The VHS spectrum is from MJD 51077
and corresponds to $\sim$2/3 of the overall maximum of the outburst. (b)
Characteristic spectra from Cyg X-3\cite{sz04}. } \label{1550_x3}
\end{figure}

Figures \ref{1550_x3}(a) and (b) compare the characteristic spectra of XTE J1550--564 and of Cyg X-3\cite{sz04}. The structure of the latter remains
uncertain. It probably contains a high-mass Wolf-Rayet star with
huge mass loss\cite{keerkwijk}, but the nature of the compact object is unknown.
We see the general similarity between the spectra of the two objects, apart from
the very strong absorption in Cyg X-3 (with the depth of the $\sim$9 keV Fe K
edge indicating the presence of a very strongly ionized Thomson-thick absorber
component). We see here an ultrasoft spectrum very similar to that of XTE
J1550--564 (as well to other ultrasoft spectra shown in Fig.\ \ref{hs}).
Interestingly, this spectrum corresponds to the highest X-ray luminosity of the
system (as, e.g., in GX 339--4). Then, there are spectra similar to the VHS one
of XTE J1550--564. Finally, we see the hardest spectra, which may be identified
with the low/hard state. However, their peak energies are at $\sim$20 keV, much
lower than the peaks at $\sim$100--200 keV characteristic to the hard state in
black-hole binaries. Thus, we cannot unambiguously classify all of the spectra
of Cyg X-3 as characteristic to black-hole binaries.

An intriguing observational feature is the form of the high-energy tail in the
ultrasoft spectra (Fig.\ \ref{hs}). It appears always to have $\Gamma\simeq 2$,
substantially harder than the tails in the VHS/IS, or HS with
stronger tails (e.g., in the soft state of Cyg X-1). Currently, there seems
to be no theoretical explanation of that $\Gamma$.

\subsection{Radiative processes in the hard state}
\label{hard}

\begin{figure}[b!]
\centerline{\includegraphics[width=6.4cm]{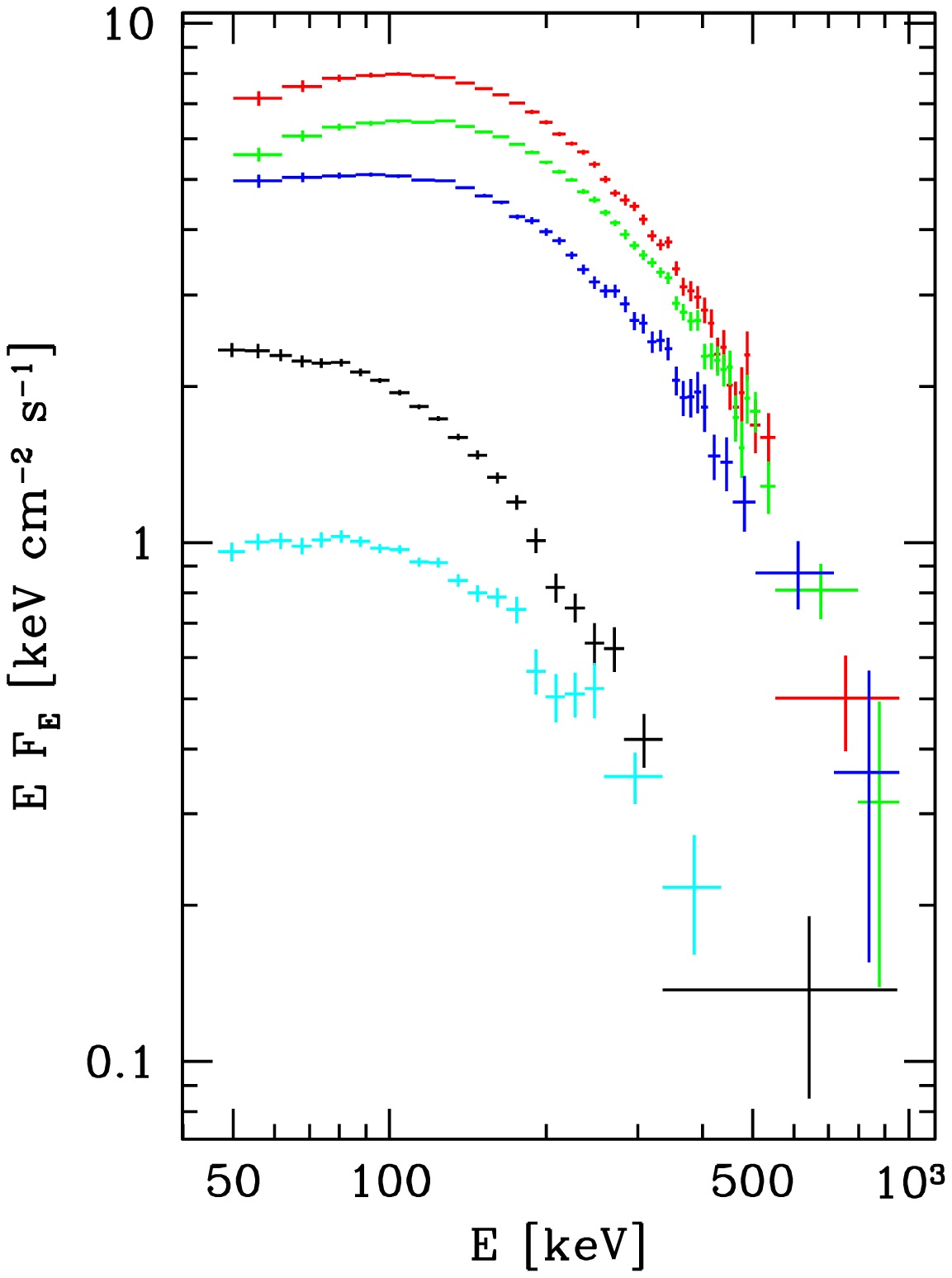}\hfill
\includegraphics[width=7.2cm]{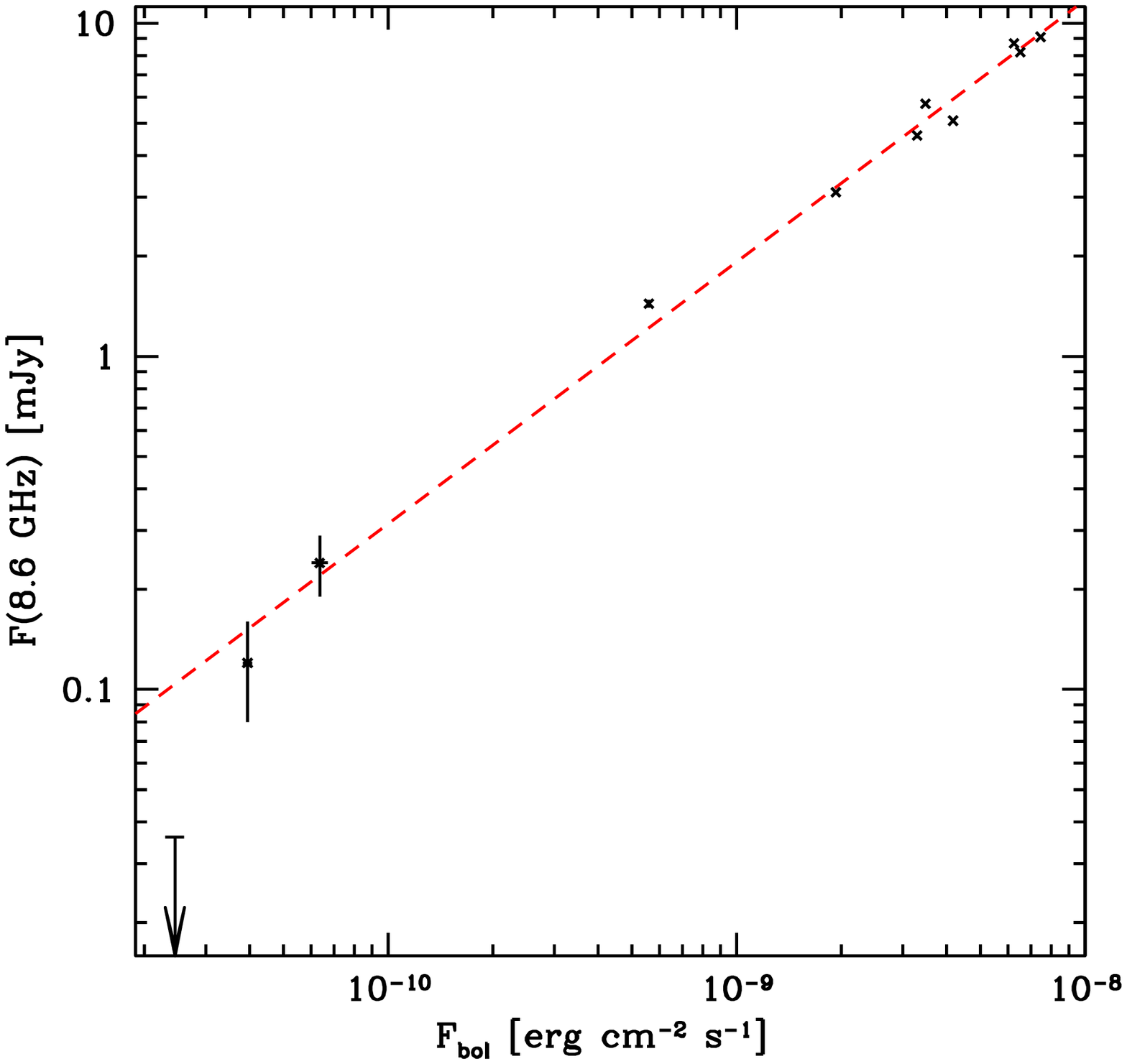}}
\caption{(a) Hard-state OSSE spectra from Cyg X-1 (3 upper spectra; red, green and blue) and GX 339--4 (2 lower spectra, black and cyan). The Cyg X-1 spectra are averages from 1991--1996 observations grouped according to the 100 keV flux (W. N. Johnson, private communication). The GX 339--4 spectra are from 1991\cite{zpm98} and 1997\cite{w02} observations. (b) Correlation between the 8.6 GHz radio flux\cite{corbel03} and the bolometric (estimated from best-fit model to PCA/HEXTE data) X\g\ flux in the hard state of GX 339--4\cite{z04}. }
\label{hard_state}
\end{figure}

As discussed above, there exists very strong evidence that the dominant
radiative process in the hard state of black-hole binaries is thermal
Comptonization. Many broad-band X\g\ spectra of black-hole binaries are very
well fitted by this model\cite{g97,f01a,f01b,w02,z02,mc02} (although a weak
non-thermal tail may be seen at $\sim$1 MeV\cite{mc02,w02}). The uniform
high-energy cutoff above $\sim$100 keV characteristic to this process is seen
in all spectra measured so far. Figure \ref{hard_state}a shows a number of
hard-state spectra from Cyg X-1 and GX 339--4, all showing the cutoff, see also
Ref.\ \citen{g98} for a number of such spectra from other objects.

The electron temperature corresponding to this cutoff is $\sim$50--100 keV.
Such temperatures are predicted for inner regions of hot accretion flows at
high accretion rates\cite{sle76,a95,ny95,e97,e98,zpm98}. Physically, the
prevalence of this temperature range is accounted for by thermostatic
properties of thermal Comptonization and \ee\ pair production\cite{mbp01,pk95}.
At higher temperatures, cooling becomes very efficient and copious pair
production starts. This reduces the available energy per electron, causing the
temperature to return to the above stable range.

Furthermore, the spectral variability patterns observed in black-hole binaries
are well interpreted by adjustment of the hot thermal plasma to variable
irradiation by soft seed photons\cite{z02,mp02,z03,z04}. For example, the hard
X-ray slope responds to variability of the soft excess in GX 339--4 in a manner
well fitted by thermal Comptonization\cite{z04}.

An alternative process proposed to account for the hard-state spectra is
synchrotron radiation from power-law electrons\cite{mff01,mncff03}. However, in
order to account for the observed X\g\ spectra, this model requires extreme
fine-tuning of both the energy and shape of the high-energy cutoff of the
electron power law\cite{z03}. Furthermore, the observed X\g\ pattern of pivoting
variability would predict extremely strong radio/IR variability\cite{z03} (due
to extrapolating to that regime of the X\g\ spectrum with a variable spectral
index), contrary to observations, showing the rms variability of both X-rays and
radio to be similar.

The main motivation for the above model is the presence of a very good
correlation between the radio and X-ray fluxes in a few bands $\simgreat$3 keV
in GX 339--4\cite{corbel03}, as well as somewhat weaker correlations in a
number of other black-hole binaries\cite{g03}. However, the presence of these
correlations does not necessarily require that both radio and X-rays are due to
the same radiative process. A likely origin of the correlation is the formation
of the radio jet out of energetic particles in the hot inner flow (i.e., the
base of the jet being that flow), which also radiates X-rays. In particular,
power-law dependencies between the jet radio flux and $\dot M$\cite{hs03} and
the accretion luminosity and $\dot M$\cite{sle76,ny95,z98} are predicted
theoretically. Thus, we  expect a corresponding power-law dependence between the
jet radio flux and the bolometric accretion $L$.

Indeed, the original correlation of the 8.6 GHz flux with band-limited
X-ray emission in GX 339--4\cite{corbel03} was later shown to hold for the
bolometric X\g\ flux from this source\cite{z04} as well (see Fig.\
\ref{hard_state}b). This correlation does not require at all that the radio and
X\g\ emission are from the same process; it can be due to the global energetics
correlations discussed above.

It seems worth to mention an analogous controversy regarding a correlation 
between the IR and X-ray emission in AGNs. An extremely good correlation was 
found between the fluxes at 3.5 $\mu$m and 2 keV in several classes of 
AGNs\cite{m84,ms82}. This was well explained by a model in which both the IR and 
X-ray emission was due to synchrotron/self-Compton emission of the same 
nonthermal electron distribution\cite{z86}, very similar to the model for the 
radio/X-ray correlation in black-hole binaries of Refs.\ \citen{mff01} and 
\citen{mncff03}. However, this model was later ruled out, e.g., after the OSSE 
showed the presence of a high-energy cutoff in Seyferts very similar to that in 
black-hole binaries\cite{g96,zpj00}. The exact origin of this correlation 
remains unknown, but it appears it is related to some global scaling relations 
of AGNs\cite{sgk04} and/or radiation of dust\cite{cvc03}.

\section{Millisecond flares from black-hole binaries}
\label{flares}

Cyg X-1 is a persistent source with rather modest variability on various
time-scales. A transition between spectral states has a typical timescale of
weeks\cite{wen01,z02} (Figure \ref{cygx1_96}), with the corresponding change in
the bolometric $L$ by a few. Shorter flaring episodes on timescales of
days\cite{cfe02} and hours\cite{sbp01,gol03} have also been observed. On
subsecond timescales, relatively weak flares or shots are seen in the
lightcurves\cite{rot74, flc99}. Typical timescales of these short events are
$\sim$10 ms, and the observed count rate increases by a factor $\simless$2.
High-frequency power spectra of Cyg X-1 show very little variability at
$\simgreat$100 Hz\cite{rgc00}.

\begin{figure}[b!]
\centerline{\includegraphics[width=6.5cm]{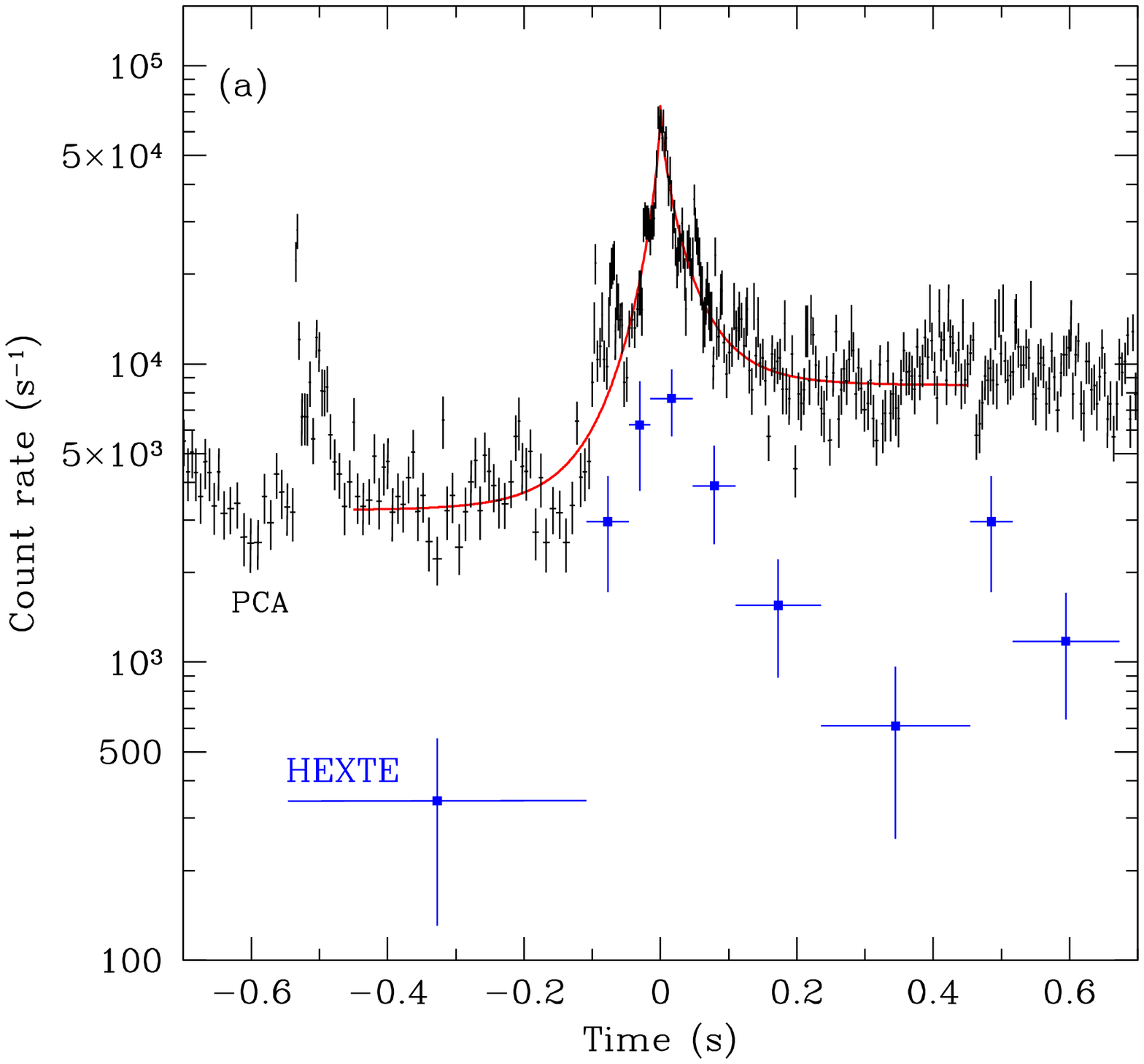}\hfill
\includegraphics[width=6.5cm]{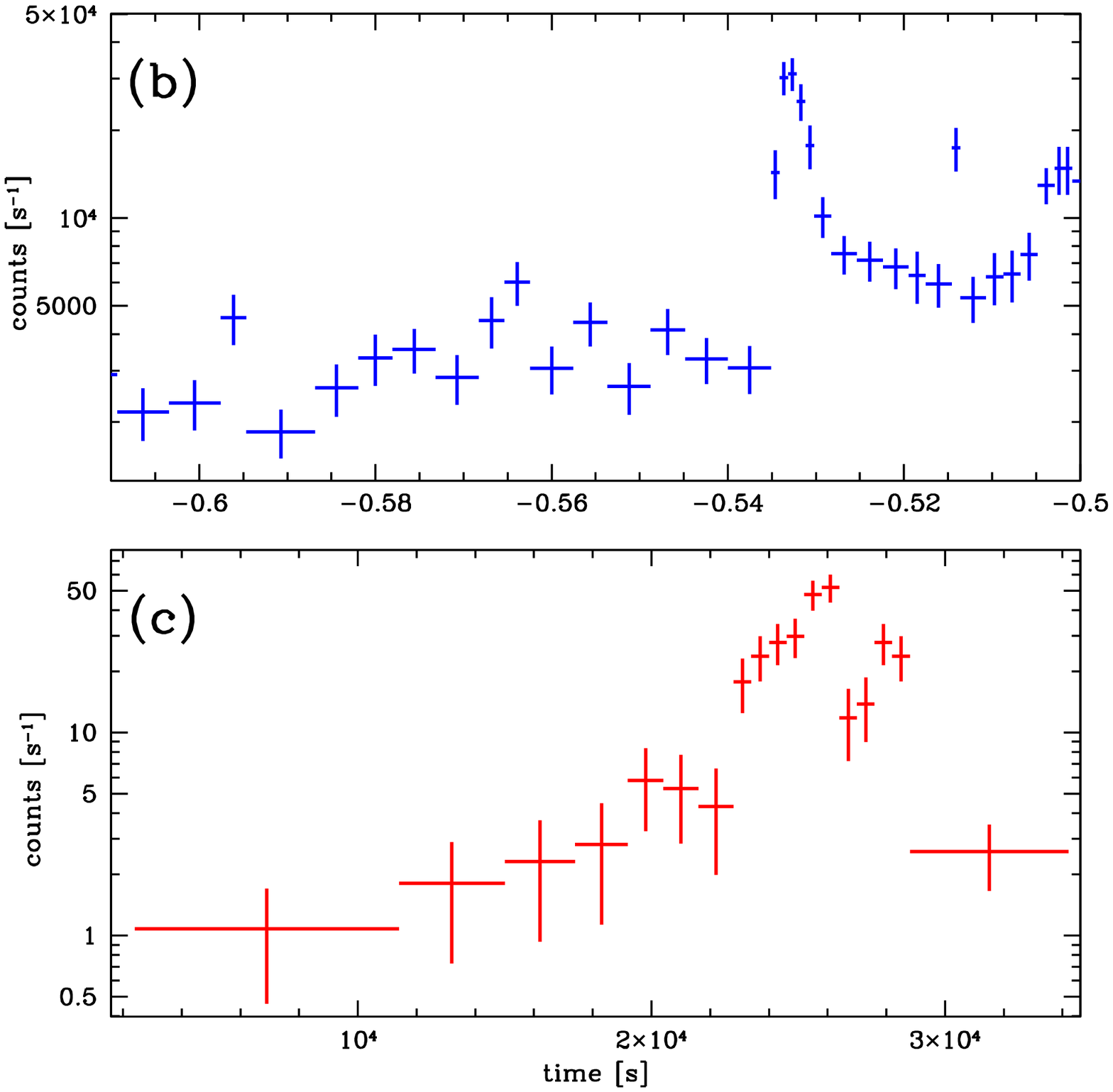}}
\caption{(a) The PCA and HEXTE (multiplied by 20 for clarity) time profiles of
the strongest millisecond flare detected from Cyg X-1 as yet\cite{gz03}. (b) A
comparison of the time profile of its precursor with (c) the 4.5--8 keV time
profile of the flare in Sgr A$^*$\cite{b01}. The vertical axes in (b) and (c)
have the same logarithmic length, and the time axis for Sgr A$^*$ is rescaled
with respect to that for Cyg X-1 by their probable mass ratio, $3\times 10^5$.
} \label{cygx1_sgrA}
\end{figure}

\begin{figure}[t!]
\centerline{\includegraphics[width=6.5cm]{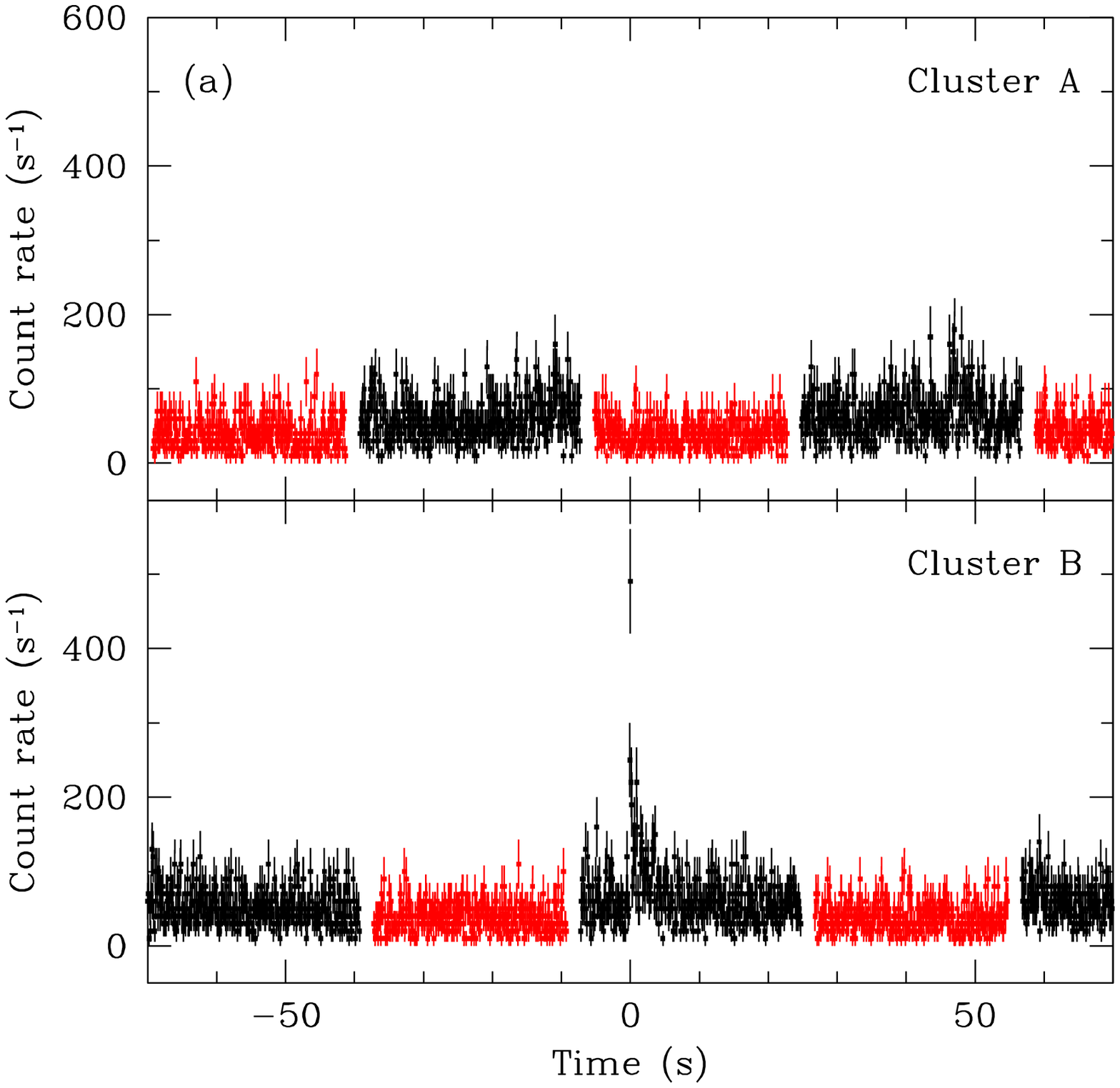}\hfill
\includegraphics[width=7.4cm]{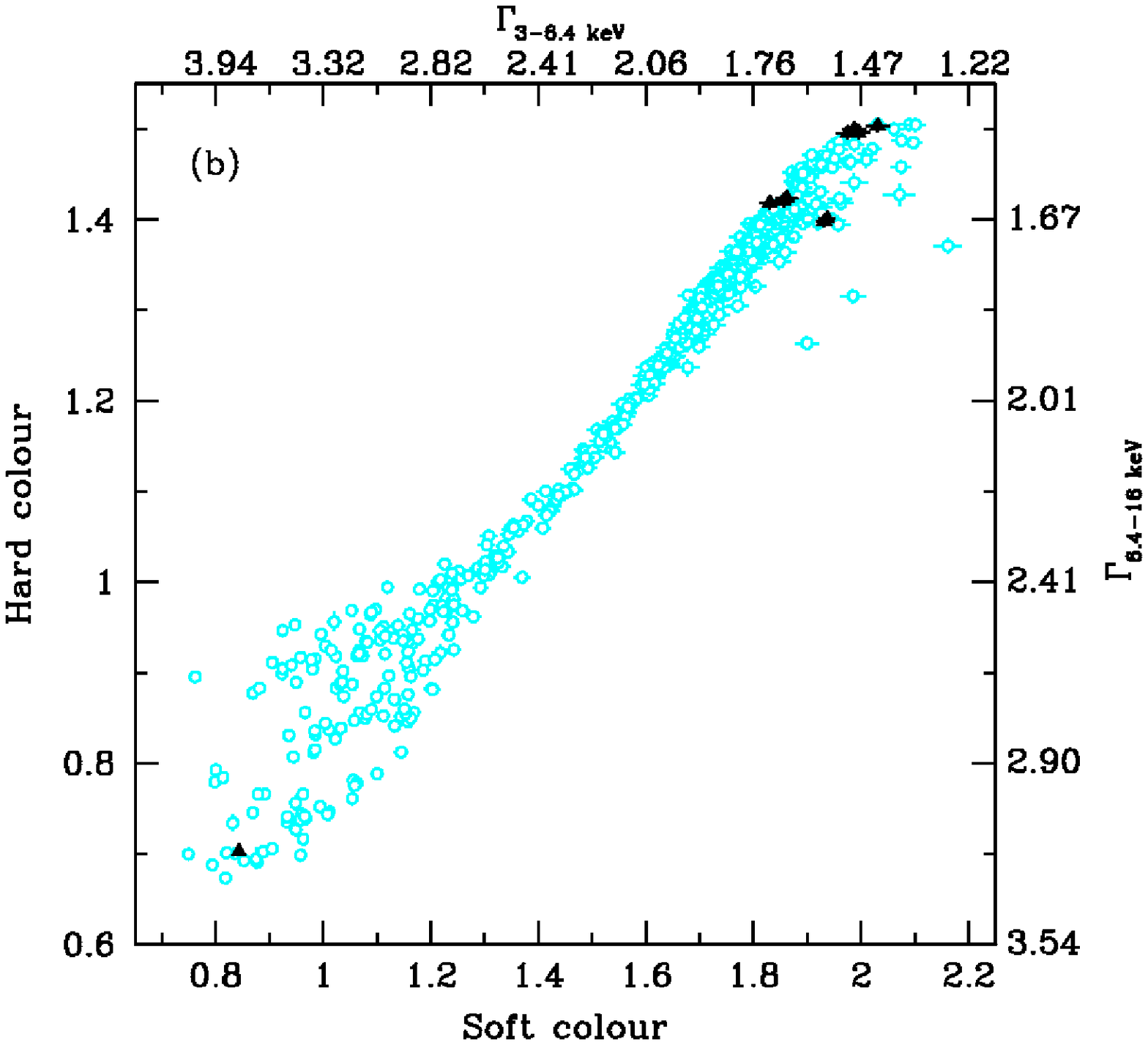}} \caption{(a) The
count rates from two HEXTE units during the strongest flare, Fig.\
\ref{cygx1_sgrA}a. The two units look alternatively at the source (black) and
the background (red; $1^\circ$ away). We see that the flare came from the
direction of Cyg X-1 within the $<1^\circ$ accuracy. (b) The positions of the
PCA observations containing the 13 brightest flares detected from Cyg
X-1\cite{gz03} (black triangles) on the color-color diagram\cite{dg03} showing
all observations available so far (cyan circles). The upper and right axis gives the average spectral index\cite{z02} in the 3--6.4 keV and 6.4--16 keV energy band, corresponding to the soft and hard color, respectively. The line corresponds to both $\Gamma$ equal.
}
\label{color_hexte}
\end{figure}

This rather modest variability of Cyg X-1 is  contrasted by the recent discovery
of powerful fast flares\cite{gz03} in the \xte/PCA lightcurves. The shortest
time-scales recorded are $\sim$1 ms and the peak count rates are up to more than
an order of magnitude above the persistent emission. Figure \ref{cygx1_sgrA}a
shows the strongest and the fastest flare found so far, during the extended soft
state of 2002. The photon energy and power spectrum no.\ 5 in Figure
\ref{cygx1_sp} correspond to the \xte\/ observation containing that flare. The
independent observation by two HEXTE units (Fig.\ \ref{color_hexte}a) confirms
the origin of the flare from within $1^\circ$ of the direction of Cyg X-1 and
rules out a background event, e.g. a high-energy particle from a solar flare or
cosmic rays. The PCA count rate profile is well fitted by a stretched
exponential with the FWHM of $\sim$25 ms (red curve in Fig.\
\ref{cygx1_sgrA}a). The 3--30 keV flux in the peak of the flare had reached
$(3.0\pm 0.6)\times 10^{-7}$ erg cm$^{-2}$ s$^{-1}$, which is $\sim$30 times the
flux before the flare. The profiles and time-scales of the shots reported in
Refs.~\citen{flc99} and \citen{nmk94} are similar to this powerful flare, but
their amplitudes are much weaker.

Even shorter time-scales were observed in the precursor $\sim$0.5 s before the
main flare, as shown in detail in Figure \ref{cygx1_sgrA}b. During the dramatic
rise the count rate increased tenfold in just 2 ms. This is of the order of the
light-travel time across an inner accretion disk around a 10$\msun$ black hole
(1 ms $\simeq 20 GM/c^3$), and it is about a half of the Keplerian period on
the minimum stable orbit in the Schwarzschild metric. It also closely
corresponds to the e-folding time of the 1999 flare\cite{b01} of Sgr A$^*$ of
$\sim$400 s, which is $\sim 30 GM/c^3$ for a $3\times 10^6\msun$ black hole. In
Figure \ref{cygx1_sgrA}c, we show the lightcurve of the Sgr A$^*$ flare
rescaled to the same dynamical timescale as Cyg X-1 flare in Fig.\
\ref{cygx1_sgrA}(b).

Apart from this fastest and most powerful flare, 12 more flares from Cyg X-1
with the peak count rate in excess of 10$\sigma$ above the preceding average
were discovered \cite{gz03}, all of them in the hard state. Fig.\
\ref{color_hexte}(b) shows the positions in the color-color diagram of
observations containing these flares. We see that the soft-state flare and
hard-state ones are both associated with the extremes of the color distribution.
More flares with smaller amplitudes can be found, and the distribution of the 12
hard-state flares is consistent with the high-amplitude tail of the log-normal
shot distribution proposed in Ref.~\citen{nm02}.

The limited photon statistics and bandwidth do not allow detailed spectral
studies of the flares, though they are sufficient to determine the overall
spectral index. The hard-state flares soften around the peak, while the powerful
soft-state flare hardens, both reaching similar $\Gamma \sim 2$ at the peak.
Still, the underlying physical mechanisms of the flares may be the same as in
the persistent emission, i.e., thermal (hard state) or nonthermal (soft state)
Comptonization (Fig.\ \ref{cygx1_comp}). The persistent emission models after
increasing $L$ and adjusting the $L_{\rm H}/L_{\rm S}$ ratio to account for
spectral softening/hardening are consistent with the flare spectra\cite{gz03}.
The models yield the peak bolometric $L\sim 0.3L_{\rm Edd}(d/2$
kpc$)^2(10$M$_\odot/M)$ for both soft and hard-state flares. This is about an
order of magnitude brighter than the persistent $L$.

\begin{wrapfigure}{1}{6.6cm}
\centerline{\includegraphics[width=7cm]{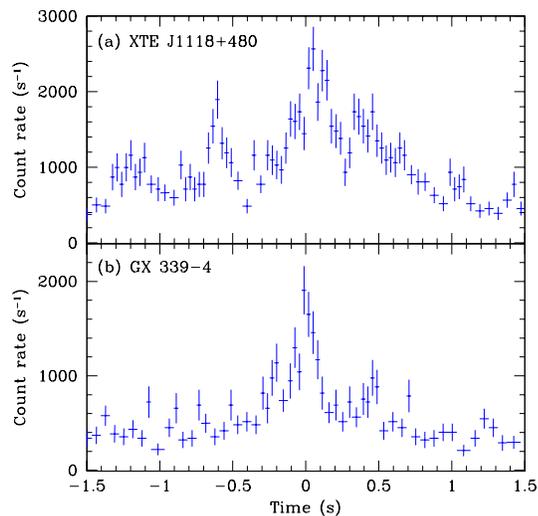}} \caption{Flares
seen from XTE J1118+480 and GX 339--4. } \label{f_1_3}
\end{wrapfigure}

Powerful millisecond flares are not inherent to Cyg X-1 only. Similar events
have been observed in two other black holes: XTE J1118+480 and GX 339--4
(Fig.\ \ref{f_1_3}). It is possible that more of the black holes exhibit
powerful flares, but limitations in the photon statistics prevent us from
detecting them.

The physical origin of the millisecond flares is yet to be established. Because
of their very short time-scales they must be due to sudden release of the
accretion energy in the inner region of the disk. This can be due to
amplification of the magnetic field in the innermost plunging region of the
disk\cite{mm03}, due to flipping among multiple state of the accretion flow
with large amplitude on short dynamical time-scales\cite{pb03} or due to
accumulation of a very large amount of energy in a flare above the disk surface
and its subsequent fast release\cite{bel99}.

\section{Discussion and conclusions}

\begin{figure}[t!]
\centerline{\includegraphics[width=11cm]{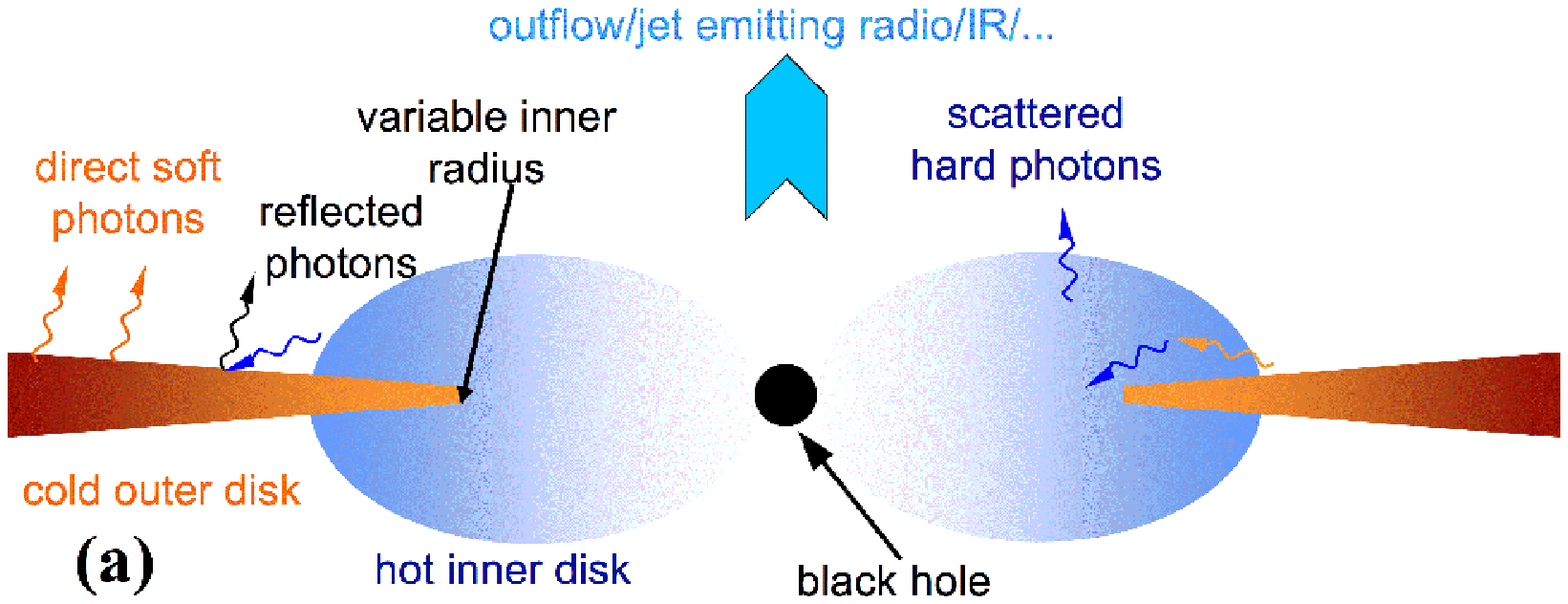}} 
%\centerline{\includegraphics[height=11.cm,angle=-90]{geo_hard.eps}}
\vskip 0.7cm
\centerline{\includegraphics[width=11cm]{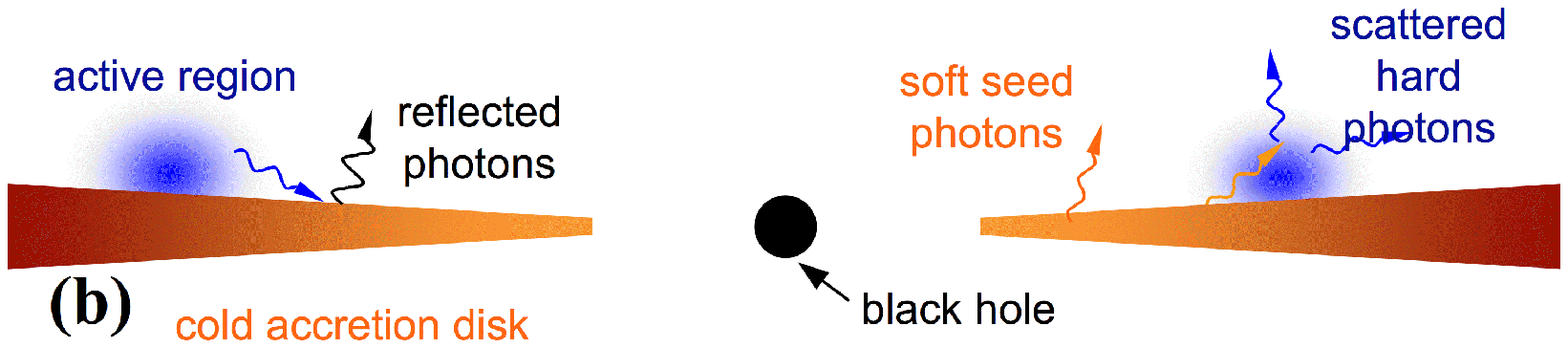}} \caption{(a) A
schematic representation of the likely geometry in the hard state,
consisting of a hot inner accretion flow surrounded by
optically-thick accretion disk. The hot flow consitutes the base
of the jet (with the counter-jet omitted from the figure for
clarity). The disk is truncated far away from the minimum stable
orbit, but it overlaps with the hot flow. The soft photons emitted
by the disk are Compton upscattered in the hot flow, and emission
from the hot flow is partly Compton-reflected from the disk. (b)
The likely geometry in the soft state consisting of flares/active
regions above an optically-thick accretion disk extending close to
the minimum stable orbit. The soft photons emitted by the disk are
Compton upscattered in the flares, and emission from the flares is
partly Compton-reflected from the disk\cite{z02}.
 }
\label{geo}
\end{figure}

We have reviewed spectral states of black hole binaries and their respective 
dominant radiative processes. In the hard state, the dominant process is thermal 
Comptonization of soft photons, most likely blackbody ones  from an outer 
accretion disk. The thermal character of the distribution of the scattering 
electrons is probably related to the most likely source geometry, consisting of 
an inner hot accretion flow\cite{a95,ny95,z98}, see Fig.\ \ref{geo}a. The 
electrons in the flow are powered by energy transfer from the hot ions, which 
receive most of the available gravitational energy. The stochastic nature of the 
energy transfer, probably via Coulomb interactions, results in the electron 
distribution being nearly thermal.

In addition, the cold medium gives rise to reprocessing, including Compton
reflection and fluorescence. Relativistic broadening observed in some cases
indicates that, in addition to an outer disk, there is some cold medium close
to the black hole, e.g, from collapse of the innermost hot flow\cite{y01}. The
variability pattern observed on long timescales indicates that the dominant
driver of the variability is changing irradiation of the hot flow by soft seed
photons from the outer cold disk\cite{z02,z03}.

In the soft states, Compton scattering is by a hybrid, thermal/nonthermal
electron distribution. The scattering plasma forms, most likely, active regions
above the surface of an accretion disk, which emit the energetically dominant
blackbody emission, see Fig.\ \ref{geo}(b). The active regions are powered by
energy transfer from inside the disk, most likely via magnetic buoyancy and
subsequent magnetic field annihilation, leading then to acceleration of
particles. This is inherently a nonthermal process, which explains the observed
nonthermal photon spectra, showing no high-energy cutoff up to high energies.
The ratio of the power in the active regions to the disk power increases in the
sequence of the soft states: ultrasoft, high, intermediate/very high. Thus, the
last state is characterized by a powerful corona surrounding most of the disk,
in a modification of the picture of Fig.\ \ref{geo}(b).

In a range of $L/\ledd$, either the hard/hot and soft/cold accretion flows are
possible. This leads to hysteresis in the long-term lightcurve of black-hole
binaries, and the corresponding long-term limit cycle.

A very interesting new phenomenon in black-hole binaries is that of millisecond
flares, extreme events occurring on timescales comparable to $GM/c^3$. In
particular, the strongest flare discovered from Cyg X-1 appears to be the most
extreme event detected from black-hole accretion yet.

\section*{Acknowledgements} This research has been supported by KBN grants 
PBZ-KBN-054/P03/2001, 5P03D00821, 2P03C00619p1,2, and 1P03D01827. We thank C. 
Done, B. Hartman, T. Maccarone and J. Poutanen for valuable discussions and/or 
comments, K. Ebisawa for the spectra of GX 339--4 from \ginga, S. Corbel for the 
GX 339--4 radio data, F. Baganoff for the Sgr A$^*$ data, and K. Beckwith for 
help with graphics. We also acknowledge the use of data obtained through the 
HEASARC online service provided by NASA/GSFC.

\end{document}